\documentclass[english,prb,floatfix,superscriptaddress,twocolumn,showpacs,amsmath,amssymb,reprint]{revtex4-1}
\usepackage[T1]{fontenc}
\usepackage[latin9]{inputenc}
\usepackage{color}
\usepackage{verbatim}
\usepackage{float}
\usepackage{amsmath}
\usepackage{amssymb}
\usepackage{graphicx}
	
\makeatletter

\providecommand{\tabularnewline}{\\}

\PassOptionsToPackage{caption=false}{subfig} 
\usepackage{hyperref}
\hypersetup{
breaklinks=true,
colorlinks=true,
citecolor=blue,
linkcolor=blue,
filecolor=blue,
urlcolor=blue
}
\IfFileExists{lmodern.sty}{\usepackage{lmodern}}{}
\makeatother

\usepackage{babel}
\begin{document}
\title{Highly-symmetric random one-dimensional spin models }
\author{V. L. Quito}
\affiliation{Department of Physics and Astronomy, Iowa State University, Ames,
Iowa 50011, USA}
\affiliation{National High Magnetic Field Laboratory, Florida State University,
Tallahassee, Florida 32306, US}
\author{P. L. S. Lopes}
\affiliation{Stewart Blusson Quantum Matter Institute, University of British Columbia,
Vancouver, British Columbia, Canada V6T 1Z4}
\affiliation{D\'{e}partement de Physique, Institut Quantique and Regroupement
Qu\'{e}b\'{e}cois sur les Mat\'{e}riaux de Pointe, Universit\'{e}
de Sherbrooke, Sherbrooke, Qu\'{e}bec, Canada J1K 2R1}
\author{Jos\'{e} A. Hoyos}
\affiliation{Instituto de F\'{i}sica de São Carlos, Universidade de S\~{a}o Paulo,
C.P. 369, S\~{a}o Carlos, SP 13560-970, Brazil}
\author{E. Miranda}
\affiliation{Gleb Wataghin Physics Institute, University of Campinas, Rua S\'{e}rgio
Buarque de Holanda, 777, CEP 13083-859 Campinas, SP, Brazil}
\date{\today}
\begin{abstract}
The interplay of disorder and interactions is a challenging topic
of condensed matter physics, where correlations are crucial and exotic
phases develop. In one spatial dimension, a particularly successful
method to analyze such problems is the strong-disorder renormalization
group (SDRG). This method, which is asymptotically exact in the limit
of large disorder, has been successfully employed in the study of
several phases of random magnetic chains. Here we develop an SDRG
scheme capable to provide in-depth information on a large class of
strongly disordered one-dimensional magnetic chains with a global
invariance under a generic continuous group. Our methodology can be
applied to any Lie-algebra valued spin Hamiltonian, in any representation.
As examples, we focus on the physically relevant cases of SO(N) and
Sp(N) magnetism, showing the existence of different randomness-dominated
phases. These phases display emergent SU(N) symmetry at low energies
and fall in two distinct classes, with meson-like or baryon-like characteristics.
Our methodology is here explained in detail and helps to shed light
on a general mechanism for symmetry emergence in disordered systems.
\end{abstract}
\maketitle

\newpage{}

\section{Introduction \label{sec:Introduction}}

Magnetism carries a historical reputation as a useful platform to
study quantum phases and transitions.\cite{sachdev2001quantum} The
convenience of magnetism does not arise from simple chance or tradition:
it comes from the easiness with which one defines symmetries and their
breaking, their accuracy to describe experimental results and the
inherent importance of quantum fluctuations. In particular, in one
spatial dimension powerful tools, which are unavailable or less potent
in higher dimensions, can be employed to gain useful insight on these
important systems. Spin chains are set apart as a truly ideal playground
in this regard.

An ingredient whose importance should not be underestimated in phase
transitions is disorder.\cite{igloi-review} Disorder is not only
intrinsic to real physical materials, playing fundamental roles in
the determination of transport properties, but it may also stabilize
distinctive phases with no analogue in clean systems. The random singlet
phase (RSP) in the disordered spin-1/2 XXZ model is the prototypical
example.\cite{fisher94-xxz} RSPs are characterized by ground states
comprised of randomly distributed and arbitrarily long singlets. They
are infinite-disorder phases, where there is a striking distinction
between the average and typical values of spin correlation functions:
while the latter decay as stretched exponentials $\sim e^{-r^{\psi}}$
with the distance $r$ between spins, the former fall off as power
laws $\sim r^{-\eta}$. The \emph{universal} tunneling exponent $\psi$
controls not only correlation functions but also thermodynamic quantities,
like the magnetic susceptibility and specific heat. In the paradigmatic
XXZ spin-1/2 chain, the tunneling exponent attains a value of $\psi=1/2$,
while $\eta=2$.\cite{fisher94-xxz}

From the statistical mechanics viewpoint (of classifying universality
classes), an infinite-randomness fixed point is an interesting concept
in its own right. In fact, infinite-randomness fixed points are much
more common than originally thought. As \emph{critical} \emph{points},
they govern a plethora of phase transitions ranging from classical
transitions in layered magnets~\cite{PhysRevB.81.144407}, passing
through quantum phase transitions in Ising magnets,~\cite{fishertransising2}
higher-spin chains,\cite{PhysRevB.66.104425,Damle2002} and quantum
rotors,~\cite{PhysRevLett.99.230601} to non-equilibrium phase transitions
in epidemic-spreading models~\cite{PhysRevLett.90.100601} (for more
examples, see, e.g., the reviews in Refs.~\onlinecite{igloi-review,*igloi-monthus-review2}).
In addition, they can occur in all spatial dimensions.~\cite{PhysRevLett.112.075702,Motrunich2000,Kovacs2011}
In contrast, there are few examples of infinite-randomness fixed points
describing \emph{stable} \emph{phases} of matter. To the best of our
knowledge, the only examples are the RSPs of the spin-1/2 XXZ and
higher-spin Heisenberg chains,~\cite{fisher94-xxz,Hyman1997,C.Monthus1997,C.Monthus1998,Refael2002,quitoprb2016}
the permutation-symmetric phases in non-Abelian anyonic chains~\cite{bonesteel-yang-prl07,fidkowski-etal-prb09},
and the so-called mesonic and bosonic RSPs in the SU(2) symmetric
spin-1 chains.~\cite{Quito_PhysRevLett.115.167201}

The stable RSPs of the spin-1/2 XXZ and spin-1 chains are particularly
noteworthy as they comprise examples of phases displaying symmetry
emergence, where the low-energy and long-wavelength physics of a system
are described by a larger symmetry than its microscopic description.
As the main result of this work, we uncover a unifying framework in
which the symmetry-enlarged infinite-randomness RSPs of the spin-1/2
XXZ chain and of the spin-1 chain are the simplest examples. The key
observation for this unification is not to look at arbitrary-spin
representations of SU(2), but rather at the fundamental vector representations
of SO($N$). We show that SO($N$)-symmetric random spin chains, in
the strong-disorder limit, realize two distinct RSPs: a meson-like
one, in which the tunneling exponent is $\psi_{M}=\frac{1}{2}$, and
a baryon-like one, with $\psi_{B}=\frac{1}{N}$ (for $N>1$). In both
cases, correlations are invariant under the larger SU$\left(N\right)$
group, with the mean correlations decaying algebraically with universal
exponent $\eta=2$. For odd $N$, there is a direct transition between
these two RSPs which is governed by an unstable SU($N$)-symmetric
infinite-randomness fixed point with baryon-like tunneling exponent
$\psi_{B}$. 

To obtain these results, we rely on the strong disorder renormalization
group (SDRG)~\cite{madasgupta,madasguptahu,bhatt-lee} (for a review,
see Refs.~\onlinecite{igloi-review,Vojtareview2006}). The SDRG method
consists in a sequential decimation of local strongly bound spins
in a chain with random exchange couplings. It is a real-space RG method
which allows one to keep track of the distributions of couplings under
coarse graining. The stronger the disorder (i.e. the larger the variance
of the distribution of coupling constants) is, the higher is the accuracy
of the method. When the fixed point is of the infinite-randomness
kind, the method is capable of capturing the corresponding long-wavelength
singular behavior exactly. In one spatial dimension, even analytic
solutions are possible. We extend here the SDRG methodology, incorporating
a general set of tools to handle arbitrary Lie groups that turns out
to be remarkably powerful. As we demonstrate, these can be used to
conveniently apply the SDRG to disordered Hamiltonians valued at any
desired Lie algebras; analytical expressions can be derived for decimation
rules and a natural basis is found for the coupling constants so that
their RG flow is maximally decoupled allowing for a simple fixed-point
analysis. This way, we see that our unifying framework is even more
general\textcolor{black}{. For concreteness, we payed p}articular
attention to the SO($N$)- and Sp($N$)-invariant Hamiltonians and
found that the Sp($N$)-invariant chains also have RSPs in their phase
diagrams. Unlike the SO($N$) chains, however, we find only meson-like
random-singlet phases. Finally, and more interestingly, we show that
the baryonic SO($N$)-symmetric RSPs and the mesonic SO($N$)- and
Sp($N$)-symmetric RSPs exhibit the previously mentioned emergent
(enlarged) SU($N$) symmetry. That is, the ground state and the low-energy
excitations are composed of SU($N$) symmetric objects.~\footnote{In general, symmetry emergence denotes a crossover in which the low-energy
long-wavelength physics of a Hamiltonian is governed by a larger symmetry
group than that governing the short-wavelength regime.~\cite{zamolodchikov89,PhysRevLett.115.166401,schmalianbatista08,Coldeaetal2010,Damle2002,Senthietal2004,Groveretal2014,batistaortiz04,fidkowski-etal-prb09,Yipetal2015,zohar2016,PhysRevB.55.8295,linetal98}} As a consequence, susceptibilities and correlation functions (or
any other observable) show emergent SU($N$) symmetry. We focus on
SO(N) and Sp(N) groups but we emphasize that Hamiltonians invariant
under any Lie group can be approached by our methods.

We would like to emphasize that SO(N) magnetism is not as exotic as
one might believe at first. Isomorphisms between algebras can be used
to relate seemingly hard to realize orthogonal symmetries to very
familiar ones. The first example is the well-known isomorphism between
so(3) and su(2), which applies to the spin-1 chains where the symmetry
emergence SU$\left(2\right)$ $\to$ SU$\left(3\right)$ was first
studied.~\cite{Quito_PhysRevLett.115.167201} The XXZ spin-1/2 chain
can be viewed as a realization of the isomorphism between $u\left(1\right)$
and $so\left(2\right)$, with symmetry emergence U$\left(1\right)$
$\to$ SU$\left(2\right)$. Another example is the algebra isomorphism
so(4)$=$su(2)$\otimes$su(2), the latter being realized in the Kugel-Khomskii
model\cite{Kugel1982} and explored in more detail here. These and
other cases reported in Ref.~\cite{QuitoLopes2017PRL} place our
present analysis as centrally relevant to many realizable systems.

This paper is organized as follows. In Sec.~\ref{sec:Physical-models-summary-results},
we discuss the broad picture of applicability of our findings. In
Sec.~\ref{sec:Lie-Groups-Tool} we summarize the necessary information
from group theory, considering our particular cases of interest, the
orthogonal and symplectic groups. This is a highly technical discussion
and readers who wish to understand the SDRG flow and its analysis
may choose to initially skip this section and return to it as seen
fit. In Sec.~\ref{sec:subgroups-1-1}, we display the SDRG decimation
rules in closed form that can, in principle, be generalized to spin
chains invariant under any Lie group rotations. In this same Section
we apply the results to SO(N) and Sp(N) symmetric Hamiltonians. In
Secs.~\ref{sec:subgroups-1-1-1}, we construct the phase diagram
of SO(N) and Sp(N) chains, using the examples of SO(4), SO(5), Sp(4)
and Sp(6). The the SO(3) case~\cite{Quito_PhysRevLett.115.167201}
also fits in the same discussion, but is not revisited here. After
that, we discuss the underlying mechanism of symmetry enhancement
in Sec.~\ref{sec:Emergent-Symmetries}, while some experimental predictions
are given in Sec.~\ref{sec:non-linear-suscep}. To contrast with
the whole discussion of the work, in Sec.~\ref{sec:When-the-emergent}
we discuss an counter-example where RSPs develop without symmetry
emergence. Finally, we summarize our finds and comment on generalizations
in Sec.~\ref{sec:conclusion}.

\section{Applicability to physical scenarios \label{sec:Physical-models-summary-results}}

Even though SO(N) and Sp(N) models look rather abstract, several specific
examples can be connected to readily known or realizable systems.
Focusing on SO(N)-invariant chains, physical scenarios can be obtained
relying on handy group isomorphisms, as we list next.

We start by listing the two cases already studied before. The first
one is the XXZ spin-1/2 chain which is well-known for its U(1) symmetry.
Its Hamiltonian is 
\begin{equation}
H=\sum_{i}J_{i}\left(S_{i}^{x}S_{i+1}^{x}+S_{i}^{y}S_{i+1}^{y}+\Delta_{i}S_{i}^{z}S_{i+1}^{z}\right),\label{eq:XXZ}
\end{equation}
where $\mathbf{S}_{i}$ are spin-1/2 operators and $J_{i}$ and $\Delta_{i}$
are coupling constant and anisotropy parameters, respectively. Due
to the isomorphism between the U(1) and SO(2) groups, the Hamiltonian~(\ref{eq:XXZ})
configures our first example of SO($N$) magnetism (with $N=2$).

For uncorrelated random couplings $J_{i}>0$ and $-\frac{1}{2}<\Delta_{i}<1$,
Fisher showed that $\Delta_{i}\rightarrow0$ under renormalization
and the corresponding (critical) phase is an RSP.~\cite{fisher94-xxz}
Thus, the corresponding fixed point is that of the random XX chain.
Even though the effective Hamiltonian does not exhibit SU(2) symmetry
(realized only when $\Delta_{i}=1$), the ground state and the corresponding
low-energy singular behavior are SU(2)-symmetric. Therefore, although
not explicitly noticed previously, this is the first example of the
SO($N$)$\rightarrow$SU($N$) symmetry-enhancement phenomenon in
random systems.

The second case comes from the SU(2) symmetric spin-1 chain, the Hamiltonian
of which is 
\begin{equation}
H=\sum J_{i}\left[\cos\theta_{i}\mathbf{S}_{i}\cdot\mathbf{S}_{i+1}+\sin\theta_{i}\left(\mathbf{S}_{i}\cdot\mathbf{S}_{i+1}\right)^{2}\right],\label{eq:S1}
\end{equation}
where $\mathbf{S}_{i}$ are spin-1 operators and $J_{i}$ and $\theta_{i}$
are parameters. Here, the isomorphism between the SO(3) and SU(2)
groups also plays a role. The SO(3) tensors can be understood, in
the SU(2) language, as quadrupolar operators constructed out of spin-1
vectors. Hence, the Hamiltonian~(\ref{eq:XXZ}) is our second example
of SO($N$) magnetism (now with $N=3$).

For uncorrelated random couplings $J_{i}$ and parameters $\theta_{i}$,
it was shown that two RSPs exist in this model and that their corresponding
fixed points do not exhibit SU(3) symmetry. However, like in the XXZ
spin-1/2 chain, the corresponding ground states and low-energy singular
behavior are SU(3) symmetric.~\cite{Quito_PhysRevLett.115.167201}
Although this was previously reported as an SU(2)$\rightarrow$SU(3)
symmetry enhancement, in this work this is just another example of
the SO($N$)$\rightarrow$SU($N$) phenomenon in random systems.

Now we report a novel example of the SO($N$)$\rightarrow$SU($N$)
symmetry-enhancement phenomenon in disordered spin chains. By analogy,
the next simplest scenario is that for $N=4$, which is identified
with the disordered version of the Kugel-Khomskii Hamiltonian.~\cite{KugelKhomskii}
In this model, each site has two orbital and two spin degrees of freedom.
With the three components of spin operators $\mathbf{S}$ and the
orbital degrees of freedom $\mathbf{T}$, it is possible to construct
the nine spin-orbital operators $S^{a}T^{b}$. The operators $\mathbf{S}$
and $\mathbf{T}$ can be chosen as the six generators of SO(4), while
the collection of SO(4) and spin-orbital operators generate the SU(4)
group. The choice of the spin and orbital vectors as generators of
SO(4) comes from the isomorphism SO(4)$\sim$SU(2)$\otimes$SU(2).
When the nine spin-orbital operators appear with the same coefficient
as the three spin and orbital operators, the model becomes SU(4)-invariant.
In general,

\begin{equation}
H=\sum_{i}J_{i}\left(\boldsymbol{S}_{i}\cdot\boldsymbol{S}_{j}+a\right)\left(\boldsymbol{T}_{i}\cdot\boldsymbol{T}_{j}+a\right),\label{eq:KK}
\end{equation}
where $J_{i}$ are random numbers taken to be positive and $a=1$
for SU(4) symmetry. If the coefficients of the spin-orbital operators
are different from the spin and orbital ones, that is, $a\ne1$, the
global SU(4) symmetry is broken down to SO(4). In the general case,
however, the Hamiltonian~(\ref{eq:KK}) configures our third example
of SO($N$) magnetism, here with $N=4$.

By constructing a phase diagram as function of $a$, this model is
found to flow under the SDRG to a nontrivial RSP exhibiting enhanced
SU(4) symmetry governed by a infinite-randomness fixed point with
$a=0$.

Another physically relevant case are the SO(6)-invariant chains. As
a consequence of the isomorphism between SO(6) and SU(4) groups, this
can be physically realized with ultra-cold alkaline-earth atoms in
the SU(4) context.~\cite{Gorshkov2010} In the presence of disorder,
a RSP can be stabilized exhibiting enlarged SU(6) symmetry. In other
words, the SU(4)-symmetric spin chain in its six-dimensional representation
(two horizontal boxes, in Young-tableau notation) is also SO(6) symmetric
and displays SU(6)-symmetric RSP physics.

Finally, SO(N) chains can also be constructed from SU(2)-symmetric
spin-$S$ chains, with $N=2S+1$. The construction is not generic,
however, as it requires fine tuning since larger degeneracies of the
multiplets of two coupled spins are required.~\cite{quitoprb2016}

\section{Lie Groups Tool Kit\label{sec:Lie-Groups-Tool}}

\textcolor{black}{The analyses of renormalization group (RG) flows
always benefit from a clever parametrization of coupling constants,
ideally one that decouples the flow of the variables as much as possible.
The SDRG, with a thermodynamically infinite number of coupling constants,
is no different. The optimal choice is determined by symmetry considerations:
a proper language keeps the covariance of the Hamiltonian under the
RG steps explicit, and is such that the decimation rules can be computed
in an as-easy-as-possible way. Such an ideal language for the SDRG
will be introduced here. It takes the form of a tool kit of Lie groups
and algebras, particularizing for our purposes to the cases of SO(N)
and Sp(N). Generally, using this tool kit one may derive the SDRG
decimation steps of spin Hamiltonians in any representation of any
group.}

\textcolor{black}{The tools in our set comprise: (i) Lie algebras
and their unique representation label scheme. (ii) Irreducible tensor
operators lying within a given representation (i.e. a generalization
of the familiar SU(2) irreducible tensor operators, naturally built
out of spherical harmonics).~\cite{quitoprb2016} (iii) The corresponding
group-invariant scalars for each representation. These scalars permit
a convenient construction of the most general group-invariant Hamiltonian.
(iv) The Wigner-Eckart theorem, which brings out the full value of
the points above in the SDRG method. It ensures independent coupling
constant RG flows for each different scalar operator that appears
in the spin Hamiltonian. This theorem, applicable to }\textcolor{black}{\emph{any}}\textcolor{black}{{}
Lie group,~\cite{Iachello_book} allows one to easily compute the
matrix elements of the irreducible tensor operators, and allows the
perturbation theory steps of the SDRG decimation to be performed easily.}

\textcolor{black}{This Section is quite mathematical, but necessary
for what follows; the general exposition here follows the conventions
of Ref.~\onlinecite{Iachello_book}.} \textcolor{black}{In order
to keep the discussion less abstract, we introduce most concepts using
the SO(N) group as a prototype, the Sp(N) case following more easily.}
The notation used throughout the paper is listed in Table~\ref{tab:notation}.

\begin{table}
\begin{centering}
\begin{tabular}{cc}
\hline 
\multicolumn{1}{|c|}{Notation} & \multicolumn{1}{c|}{Description}\tabularnewline
\hline 
\hline 
$\Upsilon$ & Irreducible representations\tabularnewline
\hline 
$\upsilon$ & Intra-representation state labels\tabularnewline
\hline 
$\mathbf{L}_{i}/\mathbf{M}_{i}/\boldsymbol{\Lambda}_{i}$ & %
\begin{tabular}{c}
SO(N)/Sp(N)/SU(N) spins on site $i$ \tabularnewline
in a given representation\tabularnewline
\end{tabular}\tabularnewline
\hline 
$T_{\upsilon}^{\Upsilon}\left(\mathbf{L}_{i}\right)$ & %
\begin{tabular}{c}
Irreducible tensor operator of rank $\ensuremath{\Upsilon}$,\tabularnewline
component $\ensuremath{\upsilon}$, as a function of $\ensuremath{\mathbf{L}_{i}}$\tabularnewline
\end{tabular}\tabularnewline
\hline 
$\mathcal{O}^{\Upsilon}\left(\mathbf{L}_{i},\mathbf{L}_{j}\right)$ & %
\begin{tabular}{c}
Scalar operator built out of tensors of rank $\Upsilon$\tabularnewline
on the link $\left(i,j\right)$ defined by corresponding spins\tabularnewline
\end{tabular}\tabularnewline
\hline 
\begin{tabular}{c}
$K^{(1)},K^{(2)}$\tabularnewline
$\bar{K}^{(1)},\bar{K}^{(2)}$\tabularnewline
\end{tabular} & SO(N)/Sp(N) coupling constants\tabularnewline
\end{tabular}
\par\end{centering}
\caption{A summary of our notation conventions. \label{tab:notation}}
\end{table}

\subsection{SO(N) group \label{sec:SO(N)-group}}

\subsubsection{Group structure and representation labeling}

The SO(N) group is defined by the set of orthogonal $\mathrm{N}\times\mathrm{N}$
matrices satisfying 

\begin{equation}
OO^{T}=O^{T}O=1,\,\,\det O=1.\label{eq:SON-def}
\end{equation}
The $O$ matrices admit an exponential description as $O=e^{iA}$,
where, from Eq.~(\ref{eq:SON-def}), $A^{T}=-A$ and $\mbox{Tr}A=0$.
Choosing a unitary representation, the anti-symmetric $A$ matrices
are Hermitian, and thus purely imaginary. A complete basis for these
matrices has $d_{SO}=\mathrm{N}\left(\mathrm{N}-1\right)/2$ elements
$L^{ab}$, the group generators.\footnote{We can assume $a<b$ if we define $L^{ab}=-L^{ba}$ whenever $a>b$. }
Therefore, one can write the $A$ matrices as linear combinations
with real coefficients $\xi_{ab}$ as

\begin{equation}
A=\sum_{a,b}\xi_{ab}L^{ab},\label{eq:SON-gen}
\end{equation}
where a normalization choice is made according to which $\mbox{Tr}\left(L^{ab}L^{cd}\right)=2\left(\delta^{ac}\delta^{bd}-\delta^{ad}\delta^{bc}\right)$.
Combining Eqs.~(\ref{eq:SON-gen}) and (\ref{eq:SON-def}), and expanding
to first order in $\xi_{ab}$, one obtains the so(N) Lie algebra

\begin{equation}
\left[L^{ab},L^{cd}\right]=i\left(\delta^{ac}L^{bd}+\delta^{bd}L^{ac}-\delta^{ad}L^{bc}-\delta^{bc}L^{ad}\right).\label{eq:SON-algebra}
\end{equation}

The distinct sets of matrices that satisfy (\ref{eq:SON-algebra})
are the representations of the algebra/group, which requires a unique
labeling scheme. SO(N) representations separate into tensor and spinor
representations. Since here we are interested only in the former,
from now on, when we talk about SO(N) representations we will be referring
to tensor representations. SO(N) representations can be denoted by
a set of integers: $\Upsilon\equiv\left[\mu_{1},\mu_{2},...,\mu_{\nu}\right]$,
with $\nu=\text{int}\left(\mathrm{N}/2\right)$. These integers are
such that, for odd N $\mu_{1}\geqslant\mu_{2}\geqslant...\geqslant\mu_{\nu}\geqslant0$
whereas for even N $\mu_{1}\geqslant\mu_{2}\geqslant...\geqslant\left|\mu_{\nu}\right|\geqslant0$.\cite{Iachello_book}
Equivalently, these $\left\{ \mu_{i}\right\} $ can be used to ascribe
a Young tableau to each representation, with each integer representing
the number of boxes in each row. As usual, anti-symmetric representations
correspond to a single column of boxes and, from the value of $\nu$,
we see that, in SO(N), anti-symmetric representations exist with at
most $\text{int}\left(\mathrm{N}/2\right)$ boxes. This is in contrast
to the familiar SU(N) case, where a column of up to $N-1$ boxes is
allowed. 

To help familiarize the reader, some examples are shown in Table \ref{tab:rep_label}.
The cases of SO(2) and SO(3) can be understood from the standard angular-momentum
physics. SO(2) representations are labeled by a single set of integers,
both positive and negative valued, $\left[\mu_{1}\right]\leftrightarrow M,\,\left|M\right|\geqslant0$,
which are nothing but the eigenvalues of the z-component of the angular-momentum
operator. This is because the Abelian group of two-dimensional rotations
in the $xy$-plane admits only one-dimensional representations with
basis vectors $\psi_{M}\left(\phi\right)=e^{iM\phi}$. As for SO(3),
the representations are again given by a single number, but now only
positive integers are allowed $\left[\mu_{1}\right]\leftrightarrow J$,
which, from standard knowledge, have a unique correspondence to the
square of the angular-momentum vector operator. The next natural example
is SO(4), but due to some caveats unique to this group, we postpone
its discussion for later. Moving on to SO(5), two integers become
necessary to identify a representation. For example, $\left[0,0\right]$
is the singlet with dimension $d_{\left[0,0\right]}=1$, $\left[1,0\right]$
is the fundamental vector (or defining) representation with dimension
$d_{\left[1,0\right]}=5$, and is represented by a single box in Young-tableau
language. Examples of anti-symmetric and symmetric representations
are, respectively $\left[1,1\right]$ and $\left[2,0\right]$, with
dimensions $d_{\left[1,1\right]}=10$ and $d_{\left[2,0\right]}=14$.
They are alternatively represented, respectively, by two vertically
and two horizontally arranged boxes in Young-tableau language. This
structure for defining anti-symmetric and symmetric representations
is general for all $\mathrm{N}$. As we move on in the text, we refer
to the representations interchangeably in the notation of $\Upsilon$,
the dimension of the representation or the corresponding Young-tableau,
as dictated by convenience.

\begin{table}[t]
\begin{centering}
\begin{tabular}{cc}
$SO\left(2\right)$ & $\left[\mu_{1}\right]\leftrightarrow M$\tabularnewline
$SO\left(3\right)$ & $\left[\mu_{1}\right]\leftrightarrow J$\tabularnewline
$SO\left(4\right)$ & $\left[\mu_{1},\mu_{2}\right],\,\mu_{1}\geqslant\left|\mu_{2}\right|\geqslant0$\tabularnewline
$SO\left(5\right)$ & $\left[\mu_{1},\mu_{2}\right],\,\mu_{1}\geqslant\mu_{2}\geqslant0$\tabularnewline
\end{tabular}
\par\end{centering}
\caption{Labeling of SO(N) representations. A set of integers must be attributed
to each case following a hierarchy as displayed. SO(2) and SO(3) cases
are well-known, from the $z$-component and the magnitude squared
of the angular momentum vector operator, respectively. For higher
N, the number of integers necessary to uniquely label representations
increases by one for every two values of N. \label{tab:rep_label}}
\end{table}

\subsubsection{SO(N) Hamiltonians and tensor operators}

With the definition of the SO(N) group and a choice of representation,
we now show how to write down the most\textcolor{red}{{} }general SO(N)-invariant
Hamiltonian. We are interested in a chain of sites, each carrying
a set of SO(N) generators $L_{i}^{ab}$ in the \emph{defining} $\left[1,0\right]$
representation of the group. We call this an SO(N) spin chain. The
question that arises is: how many objects contribute to the most general
SO(N) invariant Hamiltonian? We can gain some insight by starting
with powers of the contraction of generators. In the fundamental representation
of SO(N), the most general Hamiltonian for a pair of sites $i,j$
reads \cite{Tu_PhysRevB.78.094404}
\begin{equation}
H_{ij}=J_{ij}\left(\boldsymbol{L}_{i}\cdot\boldsymbol{L}_{j}\right)+D_{ij}\left(\boldsymbol{L}_{i}\cdot\boldsymbol{L}_{j}\right)^{2},\label{eq:pair-dot-product}
\end{equation}
where $\boldsymbol{L}_{i}\cdot\boldsymbol{L}_{j}\equiv\sum_{a<b}L_{i}^{ab}L_{j}^{ab}$.
Powers of the dot product greater than 2 are linearly dependent on
lower powers.\cite{Tu_PhysRevB.78.094404} For the third power, for
instance,
\begin{equation}
\left(\boldsymbol{L}_{i}\cdot\boldsymbol{L}_{j}\right)^{3}=\boldsymbol{L}_{i}\cdot\boldsymbol{L}_{j}+\left(1-\mathrm{N}\right)\left(\boldsymbol{L}_{i}\cdot\boldsymbol{L}_{j}\right)^{2}+\mathrm{N}-1.
\end{equation}

The SDRG method relies extensively on perturbation theory calculations
involving the projection of the spin operators on certain representations.
While Eq.~(\ref{eq:pair-dot-product}) is easy to build, projection
calculations are much more conveniently performed if one works in
the language of \emph{irreducible tensor operators}. This was first
noticed in Ref.~\onlinecite{PhysRevLett.80.4562} and extensively
applied in Ref.~\onlinecite{quitoprb2016}. For $SO(3)\sim SU(2)$
this is the language of irreducible spherical tensors of standard
quantum mechanics textbooks.\cite{Edmondsbook} Here we discuss the
general Lie group case.

General tensor operators $T_{\upsilon}^{\Upsilon}$ are defined as
satisfying~\cite{Iachello_book}

\begin{equation}
\left[L^{ab},T_{\upsilon}^{\Upsilon}\right]=\sum_{\upsilon^{'}}\left\langle \Upsilon\upsilon^{'}\left|L^{ab}\right|\Upsilon\upsilon\right\rangle T_{\upsilon^{'}}^{\Upsilon}.\label{eq:tensor_def}
\end{equation}
The labels $\Upsilon$, that previously were used to uniquely define
a representation, also specify a tensor rank. The set of labels $v$,
on the other hand, is used to uniquely specify both a state within
a given representation and a component of a tensor operator. In the
usual scenario of SO(3)$\sim$SU(2), $\upsilon$ is then the eigenvalue
of the $z$ component of the angular momentum operator which can be
chosen to be, for example, $L^{12}$. For larger $\mathrm{N}$, a
larger collection of labels is again required to specify $\upsilon$.
For us, they can be generically chosen as a set of eigenvalues of
the\emph{ Cartan subalgebra}, that is, the eigenvalues of the maximal
subset of commuting generators, i.e., that can be simultaneously diagonalized.
In general, for SO(N), the Cartan subalgebra contains $\mbox{int}\left(\mathrm{N}/2\right)$
generators. These eigenvalues are known as \emph{weights}.

\begin{figure}
\includegraphics[width=0.85\columnwidth]{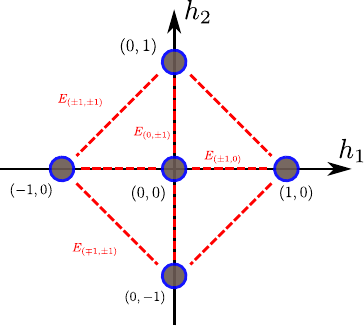}

\caption{Roots and weights for the fundamental representation of the SO(5)
group. The weights are obtained by diagonalizing the set of operators
that span the Cartan subalgebra and are used to uniquely define a
state within a given representation (in this case, the fundamental
one). In SO(5), there are $\mbox{int}\left(N/2\right)=2$ Cartan generators
and their eigenvalues, the weight vectors $\boldsymbol{\upsilon}=\left(h_{1},h_{2}\right)$
(blue circles), are two-dimensional. The $E$ operators are the root
operators and connect distinct weights (dashed red lines). \label{fig:Roots-and-weights-SO(5)}}

\end{figure}

The set of group generators can be broken down into the Cartan subalgebra
generators and the remaining set. From this remaining set,\textcolor{red}{{}
}generators can be linearly combined to produce the so-called \emph{root
operators}, that allow one to move among different weights. These
are the usual angular momentum raising and lowering operators in the
SO(3)$\sim$SU(2) case. As an example, for SO(5), the root/weight
diagram of the fundamental representation is shown in Fig.~\ref{fig:Roots-and-weights-SO(5)}.
While for a given $N$ the number of root operators is always the
same, the weights depend on the representation.

Every representation $\Upsilon$ admits a conjugate representation
with which it can be combined to build SO(N) invariant objects, or
\emph{scalar operators}, by contraction.~\cite{Iachello_book} These
scalar operators allow us to simplify the analysis of the SDRG flow
dramatically. For the SO(N) group, they are given by
\begin{equation}
\mathcal{O}^{\Upsilon}\left(\boldsymbol{L}_{i},\boldsymbol{L}_{j}\right)=\sum_{\upsilon}\left(-1\right)^{f\left(\Upsilon,\upsilon\right)}T_{\upsilon}^{\Upsilon}\left(\boldsymbol{L}_{i}\right)T_{-\upsilon}^{\Upsilon}\left(\boldsymbol{L}_{j}\right),\label{eq:CG_tensors}
\end{equation}
where $-\upsilon$ and the phase $f\left(\Upsilon,\upsilon\right)$
are fixed by the requirement that since $\mathcal{O}^{\Upsilon}$
is a scalar, i. e., it must satisfy $\left[\mathcal{O}^{\Upsilon},\boldsymbol{L}_{i}+\boldsymbol{L}_{j}\right]=\boldsymbol{0}$,
computed using (\ref{eq:tensor_def}). In SO(3), for instance, $\upsilon=M$,
$\Upsilon=J$ and $f\left(J,M\right)=M$.

\begin{figure}
\includegraphics[width=0.6\columnwidth]{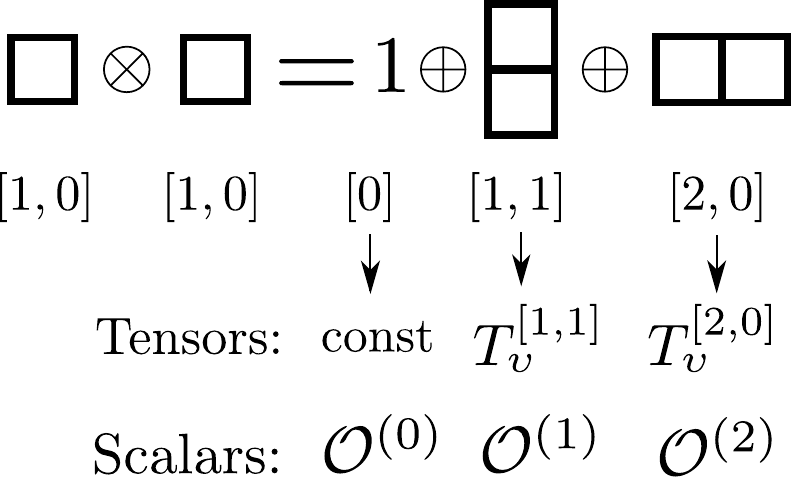}\caption{Schematic representation of how to build SO(N) scalars, starting from
the Young-tableau representation of the Clebsch-Gordan series. We
use as example the product of two fundamental representations, since
this is the relevant case for our purposes. First, associated with
each representation $\Upsilon$ coming out of the Clebsch-Gordan series
of two representations, there is a set of tensors, $T_{\upsilon}^{\Upsilon}$.
These tensors of a given rank can be contracted yielding a scalar
$\mathcal{O}^{\Upsilon}$. The most generic Hamiltonian of a pair
of sites is a linear combination of the scalars. \label{fig:scalar-construc}}
\end{figure}

Starting with tensors, one can systematically write down group-invariant
spin Hamiltonians using scalar operators as desired. To fix how many
scalars appear in a given Hamiltonian, all one needs to do is consider
the tensor product of the spins in the desired representations at
sites $i,j$. The number of terms in the Hamiltonian matches number
of terms in this Clebsch-Gordan series. For example, when two defining
representations of the SO(N) group are combined, the Clebsch-Gordan
series has three terms~\cite{FegerTables}

\begin{equation}
\left[1,\vec{0}\right]\otimes\left[1,\vec{0}\right]=\left[\vec{0}\right]\oplus\left[1,1,\vec{0}\right]\oplus\left[2,\vec{0}\right],\label{eq:3terms}
\end{equation}
where the $\vec{0}$ vectors contain as many zeroes as necessary to
complete int$\left(\mathrm{N}/2\right)$. The tensor corresponding
to $\mathcal{O}^{\left[\vec{0}\right]}$ is just a constant, and can
be neglected. We conclude, therefore, that the most general SO(N)
Hamiltonian contains two scalar operators $\mathcal{O}^{\left(1\right)}\equiv\mathcal{O}^{\left[1,1\right]}$
and $\mathcal{O}^{\left(2\right)}\equiv\mathcal{O}^{\left[2,0\right]}$,
the same number of terms as discussed before in Eq.~(\ref{eq:pair-dot-product}).
Figure~\ref{fig:scalar-construc} displays the schematic association
of the lowest representations, tensors and scalars, as well as the
Young-tableau notation of SO(N).

Having determined all the scalar operators, the most general group-invariant
Hamiltonian is written as an arbitrary linear combination of them.
Restricting interactions to first neighbors only and neglecting constant
terms, the Hamiltonian reads
\begin{align}
H & =\sum_{i}H_{i,i+1}\nonumber \\
H_{i,j} & =\sum_{\Upsilon}K^{\Upsilon}\mathcal{O}^{\Upsilon}\left(\boldsymbol{L}_{i},\boldsymbol{L}_{j}\right).
\end{align}
To make the notation clearer, we replace $\Upsilon=\left[0,0\right],\,\left[1,1\right],\,\left[2,0\right]$
by $i=0,1,2$, respectively, as labels. Particularizing to SO(N) in
its fundamental representation, only two terms contribute,

\begin{equation}
H_{i,i+1}=K_{i}^{\left(1\right)}\mathcal{O}^{\left(1\right)}\left(\boldsymbol{L}_{i},\boldsymbol{L}_{i+1}\right)+K_{i}^{\left(2\right)}\mathcal{O}^{\left(2\right)}\left(\boldsymbol{L}_{i},\boldsymbol{L}_{i+1}\right).\label{eq:Hpair_tensors}
\end{equation}

Using Eq~(\ref{eq:tensor_def}) to determine the exact structure
of the tensor operators and their corresponding scalars in terms of
spins $\mathbf{L}_{i}$ is computationally tedious and demanding.
While $\mathcal{O}^{\left(1\right)}\left(\boldsymbol{L}_{i},\boldsymbol{L}_{i+1}\right)=\boldsymbol{L}_{i}\cdot\boldsymbol{L}_{i+1}$
has a form reminiscent of the Heisenberg Hamiltonian for any group,
$\mathcal{O}^{\left(2\right)}\left(\boldsymbol{L}_{i},\boldsymbol{L}_{i+1}\right)$
can have a more complicated (non-bilinear) structure. For SO(N) (and
Sp(N) below), however, a fortunate shortcut exists. The trick is to
take advantage of the SU(N) group, of which both SO(N) and Sp(N) are
subgroups. 

The SU(N) group has $\mathrm{N}^{2}-1$ generators we will call $\Lambda^{\mu}$.
As discussed, the lowest order non-trivial SU(N) scalar operator is
$\mathcal{O}^{\left(1\right)}$ \cite{PhysRevB.70.180401}
\begin{equation}
H_{i,j}^{SU(\mathrm{N})}=\sum_{\mu=1}^{\mathrm{N}^{2}-1}\Lambda_{i}^{\mu}\Lambda_{j}^{\mu}=\boldsymbol{\Lambda}_{i}\cdot\boldsymbol{\Lambda}_{j}.\label{eq:SUN_Hamilt}
\end{equation}
As the unitary algebra contains the orthogonal one, su(N)$\supset$so(N),
we can focus on the $\mathrm{N}\left(\mathrm{N}-1\right)/2$ purely
imaginary generators of SU(N) which are, in fact, the $d_{SO}$ generators
of the SO(N) group $\left(\Lambda_{i}^{\mu}=L_{i}^{ab}\right)$. Labeling
these with $\mu=1,\ldots,d_{SO}$, this simple observation allows
us to write
\begin{equation}
\mathcal{O}^{\left(1\right)}\left(\boldsymbol{L}_{i},\boldsymbol{L}_{j}\right)=\boldsymbol{L}_{i}\cdot\boldsymbol{L}_{j}\equiv\sum_{\mu=1}^{d_{SO}}\Lambda_{i}^{\mu}\Lambda_{j}^{\mu}.\label{eq:O1}
\end{equation}
The SU(N) invariant Hamiltonian (\ref{eq:SUN_Hamilt}) must also be
SO(N) invariant and, since $\mathcal{O}^{\left(1\right)}$ was built
out of the $\mathrm{N}\left(\mathrm{N}-1\right)/2$ purely imaginary
generators, the remaining terms in it must immediately give us $\mathcal{O}^{\left(2\right)}$.
Indeed, $\mathcal{O}^{\left(2\right)}$ can be decomposed as 

\begin{align}
\mathcal{O}^{\left(2\right)}\left(\boldsymbol{L}_{i},\boldsymbol{L}_{j}\right) & =\sum_{\mu=d_{SO}+1}^{\mathrm{N}^{2}-1}\Lambda_{i}^{\mu}\Lambda_{j}^{\mu}\nonumber \\
 & =\boldsymbol{L}_{i}\cdot\boldsymbol{L}_{j}+\frac{2}{\mathrm{N}-2}\left(\boldsymbol{L}_{i}\cdot\boldsymbol{L}_{j}\right)^{2},\label{eq:O2}
\end{align}
where the coefficients can be found by direct computation~\footnote{One way of computing the coefficient of the quadratic term is by adding
Eqs.~(\ref{eq:O1}) and (\ref{eq:O2}), which makes the resulting
object SU(N) invariant. By summing Eqs. 7 and 8 of Ref.~\cite{Tu_PhysRevB.78.094404},
one obtains the coefficient of the quadratic term.}. Computationally, working with SU(N) matrices is a much easier task
than the complete determination of $\mathcal{O}^{\left(2\right)}$
by means of the route given in Fig.~\ref{fig:scalar-construc} and
Eqs.~(\ref{eq:tensor_def}) and (\ref{eq:CG_tensors}). This process
provides us with the Hamiltonian in terms of $\mathcal{O}^{\left(1\right)}$
and $\mathcal{O}^{\left(2\right)}$ without ever writing the tensor
operators explicitly.

Allowing for disorder to define site-dependent couplings, the disordered
SO(N) invariant Hamiltonian becomes

\begin{equation}
H=\sum_{i}\left[K_{i}^{\left(1\right)}\mathcal{O}^{\left(1\right)}\left(\boldsymbol{L}_{i},\boldsymbol{L}_{i+1}\right)+K_{i}^{\left(2\right)}\mathcal{O}^{\left(2\right)}\left(\boldsymbol{L}_{i},\boldsymbol{L}_{i+1}\right)\right],\label{eq:HamiltSON}
\end{equation}
with $\mathcal{O}^{\left(1\right)}$ and $\mathcal{O}^{\left(2\right)}$
given in Eqs.~(\ref{eq:O1}) and (\ref{eq:O2}), and $K_{i}^{\left(1\right)}=J_{i}-\frac{\mathrm{N}-2}{2}D_{i}$,
$K_{i}^{\left(2\right)}=\frac{\mathrm{N}-2}{2}D_{i}$ {[}if one wishes
to compare with the notation Eq.~(\ref{eq:pair-dot-product}){]}.
As a final remark, note that the most general SO($N$)-symmetric spin
chain can be recast as a special anisotropic SU$\left(N\right)$ spin
chain. This realization is very useful when analyzing the RG flow.

\subsection{Sp(N)}

Much of the previous analysis is in fact group and algebra independent,
so we can be more concise for the Sp(N) case. We restrict ourselves
to introducing the group, making a few comments, and moving straight
to the most general Hamiltonian, which can be found in a similar procedure
as described above. A general element of Sp(N), where N is assumed
to be even, satisfies the symplectic relation

\begin{equation}
U^{T}JU=J,\,\,\,\,J=\left(\begin{array}{cc}
0 & I\\
-I & 0
\end{array}\right),
\end{equation}
where $I$ is the $\mathrm{N}/2$-dimensional identity matrix. Writing
$U=\exp\left(i\theta\cdot\mathbf{M}\right)$, and expanding to first
order in $\mathbf{M}$, we find

\begin{eqnarray}
\left(\theta\cdot\mathbf{M}\right)^{T}J+J\left(\theta\cdot\mathbf{M}\right) & = & 0.
\end{eqnarray}

Following a similar reasoning as implemented for SO(N), we are able
to build tensor operators. First, the scalar $\mathcal{O}^{\left(1\right)}$
is constructed by contracting all the $d_{Sp}=\mathrm{N}\left(\mathrm{N}+1\right)/2$
Sp(N) generators. These generators can be identified with some SU(N)
operators, which is guaranteed by Sp(N)$\subset$SU(N). Analogously
to the SO(N) case, $\mathcal{O}^{\left(2\right)}$ can be constructed
with the remaining $\mathrm{N}\left(\mathrm{N}-1\right)/2-1$ SU(N)
generators. Thus,

\begin{align}
\mathcal{H} & =\sum_{i}\left(K_{i}^{\left(1\right)}\sum_{\mu=1}^{d_{Sp}}\Lambda_{i}^{\mu}\Lambda_{i+1}^{\mu}+K_{i}^{\left(2\right)}\sum_{\mu=d_{Sp}+1}^{\mathrm{N}^{2}-1}\Lambda_{i}^{\mu}\Lambda_{i+1}^{\mu}\right)\nonumber \\
 & =\sum_{i}\left(K_{i}^{\left(1\right)}\mathcal{O}^{\left(1\right)}\left(\boldsymbol{M}_{i},\boldsymbol{M}_{i+1}\right)+K_{i}^{\left(2\right)}\mathcal{O}^{\left(2\right)}\left(\boldsymbol{M}_{i},\boldsymbol{M}_{i+1}\right)\right).\label{eq:AnisSpN}
\end{align}
Writing the explicit form of the scalar operators, we have
\begin{align}
\mathcal{O}^{\left(1\right)}\left(\boldsymbol{M}_{i},\boldsymbol{M}_{j}\right) & =\boldsymbol{M}_{i}\cdot\boldsymbol{M}_{j}\\
\mathcal{O}^{\left(2\right)}\left(\boldsymbol{M}_{i},\boldsymbol{M}_{j}\right) & =\boldsymbol{M}_{i}\cdot\boldsymbol{M}_{j}\nonumber \\
 & +\frac{1}{\frac{\mathrm{N}}{2}+1}\left(\boldsymbol{M}_{i}\cdot\boldsymbol{M}_{j}\right)^{2}-\frac{2\left(\mathrm{N}+1\right)}{\mathrm{N}\left(\frac{\mathrm{N}}{2}+1\right)}.
\end{align}

\subsection{Generalized Wigner-Eckart Theorem \label{sec:subgroups-1}}

We now provide the final ingredient that, using the information above,
allows one to derive SDRG rules: the Wigner-Eckart theorem. One may
be familiar with this theorem from applications of group theory to
selection rules for angular momentum, the particular case of the SU(2)$\sim$SO(3)
group. Here, we remind the reader that the theorem is valid for \emph{any}
Lie group and reads~\cite{Iachello_book}
\begin{equation}
\left\langle \Upsilon_{1}\upsilon_{1}\left|T_{\upsilon}^{\Upsilon}\right|\Upsilon_{2}\upsilon_{2}\right\rangle =\left\langle \Upsilon_{1}\upsilon_{1}|\Upsilon\upsilon\Upsilon_{2}\upsilon_{2}\right\rangle \left\langle \Upsilon_{1}\left\Vert T^{\Upsilon}\right\Vert \Upsilon_{2}\right\rangle .\label{eq:WE-theorem}
\end{equation}
Again the matrix elements $\left\langle \Upsilon_{1}\lambda_{1}|\Upsilon\lambda\Upsilon_{2}\lambda_{2}\right\rangle $
are Clebsch-Gordan coefficients connecting the basis of the ``added
representation'' (analogous to total angular momentum) with the basis
that is formed by the tensor product of two representations (say,
from distinct sites). The reduced matrix elements $\left\langle \Upsilon_{1}\left\Vert T^{\Upsilon}\right\Vert \Upsilon_{2}\right\rangle $
are \emph{independent of} $\upsilon_{1},\upsilon_{2}$, being constant
for given $\Upsilon,$ $\Upsilon_{1}$ and $\Upsilon_{2}$. This factorization,
in terms of Clebsch-Gordan coefficients and reduced matrix elements,
simplifies the SDRG analysis dramatically and justifies \emph{a posteriori}
the introduction of these objects.\footnote{Sometimes, a multiplicity label $\beta$ is included to account for
possible degeneracies coming from the Clebsch-Gordan series. Such
distinction is not important for our purposes here. }

\section{Strong-disorder RG decimation rules \label{sec:subgroups-1-1}}

In this section we derive the SDRG decimation rules for any spin system
invariant under global Lie group transformations. For concreteness,
in Secs.~\ref{sec:RG-Steps-closed-form} and \ref{sec:RG-Steps-closed-form-Sp(N)}
we particularize our calculations for the SO($N$)- and Sp($N$)-symmetric
spin chains the corresponding Hamiltonians of which are (\ref{eq:HamiltSON})
and (\ref{eq:AnisSpN}), respectively. We emphasize that the process
here described is general in the sense that it is independent of the
representation and the number of scalar operators at each link.

The SDRG procedure starts by probing the chain for the most strongly
coupled pair of spins. This is determined by the pair with the largest
gap between ground and first excited multiplets (which we call the
``local gap''). The assumed large variance of the distribution of
couplings implies that, with high probability, its neighboring links
are much weaker. We thus focus on the 4-site problem

\begin{equation}
\mathcal{H}=\mathcal{H}_{1,2}+\mathcal{H}_{2,3}+\mathcal{H}_{3,4}\ ,
\end{equation}
assuming that $\left(2,3\right)$ is the strongly coupled pair and
considering $\mathcal{H}_{1,2}+\mathcal{H}_{3,4}$ as a small perturbation
to $\mathcal{H}_{2,3}$\@. As we saw in Section~\ref{sec:Lie-Groups-Tool},
we can write

\begin{eqnarray}
\mathcal{H}_{2,3} & = & \sum_{\Upsilon}K_{2}^{\Upsilon}\mathcal{O}^{\Upsilon}\left(\mathbf{L}_{2},\mathbf{L}_{3}\right).\label{eq:init_Ham}
\end{eqnarray}
In this Section, with some abuse of notation, $\mathbf{L}_{i}$ stands
for operators of \emph{any} representation \emph{any} Lie group {[}or
keep in mind SO(N) \emph{or} Sp(N) for concreteness{]}. Also, as described
ahead, at initial RG steps $\mathbf{L}_{i}$ corresponds to the generators
of the defining representation of the group, but as the RG proceeds,
that need not be the case. 

\begin{figure}
\includegraphics[width=1\columnwidth]{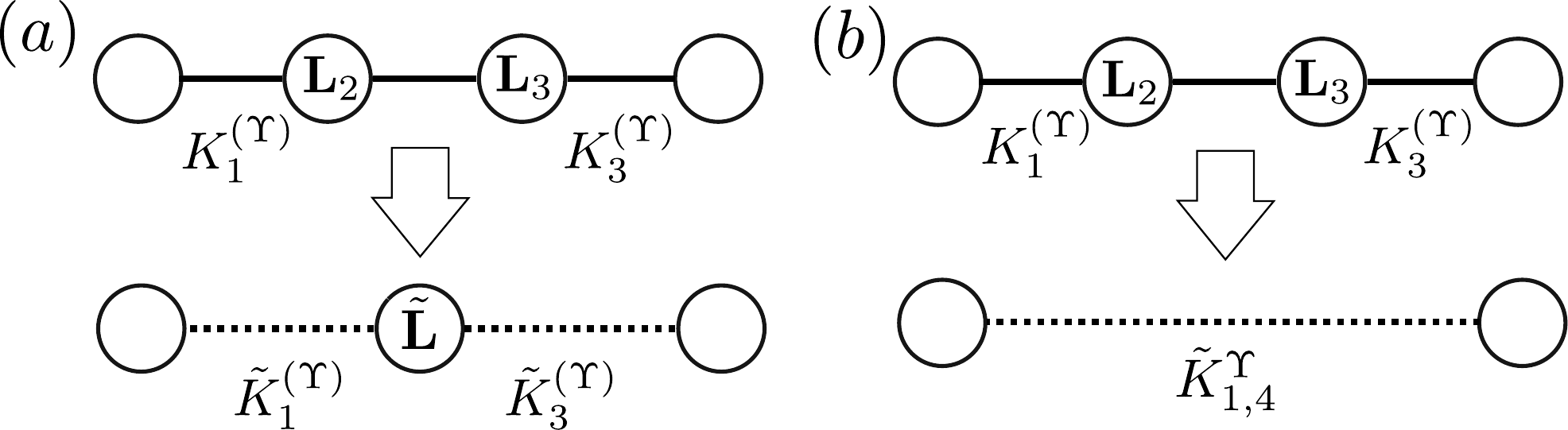}

\caption{Schematic representation of the SDRG decimation steps. Depending on
whether the local ground multiplet of the spin pair on sites 2 and
3 is degenerate or not, decimations will follow respectively from
first- {[}panel (a){]} or second-order {[}panel (b){]} perturbation
theory. The ground multiplet is fixed by the values of $K_{2}^{\Upsilon}$,
the couplings corresponding to tensors of rank $\Upsilon$. Couplings
of different ranks are not mixed by the decimation.\label{fig:Schematic-decimation}}
\end{figure}

The crucial information to be obtained from (\ref{eq:init_Ham}) is
its ground multiplet, which depends on the set of constants $\left\{ K_{2}^{\Upsilon}\right\} $.
The RG decimations project the Hamiltonian of each strongly bound
pair of spins onto such ground multiplet, and two distinct classes
of situations arise: either these representations are singlets (one-dimensional)
or they transform as some other higher-dimensional representation.
The SDRG will allow the group representations and coupling constants
to flow according to which of these two cases happens for each pair
of strongly bound spins, as we explain in the next sub-sections. A
pictorial representation of these two possible decimation steps is
shown in Fig.~\ref{fig:Schematic-decimation}. 

\subsection{First-order perturbation theory \label{subsec:First-order-perturbation-theory-gen}}

Let us assume that the energy of the strongly coupled sites 2 and
3 is great enough to justify freezing them in their two-spin ground
multiplet. If the coupling constants $\left\{ K_{2}^{\Upsilon}\right\} $
are such that this ground multiplet is degenerate, then the 4-site
problem can be treated perturbatively as an effective 3-site problem.
The middle site then corresponds to a spin-object corresponding to
the ground state manifold of the previous $\left(2,3\right)$ link
(see Fig.~\ref{fig:Schematic-decimation}). Accordingly, its couplings
to sites 1 and 4 (namely $\tilde{K}_{1}^{\Upsilon}$ and $\tilde{K}_{3}^{\Upsilon}$)
receive corrections to first-order in perturbation theory.

By symmetry, the ground multiplet of $\left(2,3\right)$ will transform
as an irreducible representation $\tilde{\Upsilon}$ of the group.
The generators in that representation (the ``new spin'' operators)
will be denoted by $\tilde{\mathbf{L}}$. Since the SDRG must preserve
the global symmetry, the effective Hamiltonian reads 

\textbf{
\begin{eqnarray}
\tilde{\mathcal{H}} & = & \sum_{\Upsilon}\tilde{K}_{1}^{\Upsilon}\mathcal{O}^{\Upsilon}\left(\mathbf{L}_{1},\tilde{\mathbf{L}}\right)+\sum_{\Lambda}\tilde{K}_{3}^{\Upsilon}\mathcal{O}^{\Upsilon}\left(\tilde{\mathbf{L}},\mathbf{L}_{4}\right).\label{eq:effec-first-order}
\end{eqnarray}
}The challenge here is to find the renormalized couplings $\tilde{K}_{i}^{\Upsilon}$.
According to (\ref{eq:CG_tensors}), we need the matrix elements of
$T_{\lambda}^{\Upsilon}\left(\mathbf{L}_{i}\right)$ ($i=2,3$) within
the $\tilde{\Upsilon}$ space. Using the Wigner-Eckart theorem in
Eq.~(\ref{eq:WE-theorem}) twice, for both the matrix elements of
$T_{\upsilon}^{\Upsilon}\left(\mathbf{L}_{i}\right)$ ($i=2,3$) and
$T_{\upsilon}^{\Upsilon}\left(\tilde{\mathbf{L}}\right)$, one finds

\begin{eqnarray}
 &  & \left\langle \tilde{\Upsilon}\upsilon_{1}\left|T_{\upsilon}^{\Upsilon}\left(\mathbf{L}_{i}\right)\right|\tilde{\Upsilon}\upsilon_{2}\right\rangle \label{eq:CG-1}\\
 & = & \left\langle \tilde{\Upsilon}\upsilon_{1}|\Upsilon\upsilon\tilde{\Upsilon}\upsilon_{2}\right\rangle \left\langle \Upsilon_{2}\Upsilon_{3};\tilde{\Upsilon}\left\Vert T^{\Upsilon}\left(\mathbf{L}_{i}\right)\right\Vert \Upsilon_{2}\Upsilon_{3};\tilde{\Upsilon}\right\rangle ,\nonumber \\
 &  & \left\langle \tilde{\Upsilon}\upsilon_{1}\left|T_{\upsilon}^{\Upsilon}\left(\tilde{\mathbf{L}}\right)\right|\tilde{\Upsilon}\upsilon_{2}\right\rangle \label{eq:CG-2}\\
 & = & \left\langle \tilde{\Upsilon}\upsilon_{1}|\Upsilon\upsilon\tilde{\Upsilon}\upsilon_{2}\right\rangle \left\langle \Upsilon_{2}\Upsilon_{3};\tilde{\Upsilon}\left\Vert T^{\Upsilon}\left(\tilde{\mathbf{L}}\right)\right\Vert \Upsilon_{2}\Upsilon_{3};\tilde{\Upsilon}\right\rangle .\nonumber 
\end{eqnarray}
As long as the Clebsch-Gordan coefficients are non-zero, we can divide
these equations, and use the fact we are within the subspace of fixed
$\tilde{\Upsilon}$ (i.e., the local ground multiplet), to obtain
the following operator identity within the $\tilde{\Upsilon}$ representation 

\begin{eqnarray}
T_{\upsilon}^{\Upsilon}\left(\mathbf{L}_{i}\right) & = & \frac{\left\langle \Upsilon_{2}\Upsilon_{3};\tilde{\Upsilon}\left\Vert T^{\Upsilon}\left(\mathbf{L}_{i}\right)\right\Vert \Upsilon_{2}\Upsilon_{3};\tilde{\Upsilon}\right\rangle }{\left\langle \Upsilon_{2}\Upsilon_{3};\tilde{\Upsilon}\left\Vert T^{\Upsilon}\left(\tilde{\mathbf{L}}\right)\right\Vert \Upsilon_{2}\Upsilon_{3};\tilde{\Upsilon}\right\rangle }T_{\upsilon}^{\Upsilon}\left(\tilde{\mathbf{L}}\right)\\
 & \equiv & \beta_{i}\left(\Upsilon,\tilde{\Upsilon},\Upsilon_{2},\Upsilon_{3}\right)T_{\upsilon}^{\Upsilon}\left(\tilde{\mathbf{L}}\right)\,\,\,i=2,3.
\end{eqnarray}

Comparing with Eq.~(\ref{eq:effec-first-order}), the couplings are
corrected by

\begin{eqnarray}
\tilde{K}_{i}^{\left(\Upsilon\right)} & = & \beta_{i}\left(\Upsilon,\tilde{\Upsilon},\Upsilon_{2},\Upsilon_{3}\right)K_{i}^{\left(\Upsilon\right)},\,\,\,\,i=1,3,\label{eq:K-order-1}
\end{eqnarray}
controlled uniquely by the reduced matrix elements of the tensor operators
within the ground state multiplet representation. The value of the
Wigner-Eckart theorem cannot be overstated here: it guarantees that
the renormalized Hamiltonian both remains written in terms of only
scalar operators and that distinct ranks are not mixed. 

First order perturbation theory fails whenever the right hand sides
of Eqs.~(\ref{eq:CG-1}) and (\ref{eq:CG-2}) vanish. For concreteness,
focus on Eq.~(\ref{eq:CG-1}). Two cases arise for which the coefficient
vanishes:

(i) When $\tilde{\Upsilon}\notin\Upsilon\otimes\tilde{\Upsilon}$.
In this case the very Clebsch-Gordan coefficients $\left\langle \tilde{\Upsilon}\upsilon_{1}|\Upsilon\upsilon\tilde{\Upsilon}\upsilon_{2}\right\rangle $
vanish. This case is the easiest to predict, since it comes directly
from the Clebsch-Gordan series of the group and does not rely on dynamics.

(ii) When $\tilde{\Upsilon}\in\tilde{\Upsilon}\otimes\Upsilon$, but
$\left\langle \Upsilon_{2}\Upsilon_{3};\tilde{\Upsilon}\left\Vert T^{\Upsilon}\left(\mathbf{L}_{i}\right)\right\Vert \Upsilon_{2}\Upsilon_{3};\tilde{\Upsilon}\right\rangle $
is still zero. This is a more exotic scenario, but is present even
in the more familiar SU(2) problem.~\cite{quitoprb2016} Since there
is no (easy) way to predict when this happens \emph{a priori}, one
has to compute such reduced matrix element explicitly to find out
whether it is zero or not.

Case (i) happens whenever the ground state is a singlet. This is a
natural situation and leads us to deal with the problem within second-order
perturbation theory, as explained in the next sub-section. In any
other situation in which one of the cases listed above happens, the
neighboring couplings are immediately renormalized to zero and the
SDRG flow, as derived here, becomes pathological. Dealing with this
situation would require going to the next order in perturbation theory
and in general, as exemplified in Ref.~\onlinecite{quitoprb2016},
the form of the Hamiltonian is not maintained. This complicates considerably
the analysis. As we show later, case (i) happens in a region of the
Sp(N) anti-ferromagnetic phase diagram. The SO(N) Hamiltonian, however,
is protected against such anomalies by the location of SU(N)-symmetric
points in its phase diagram. This will become evident in next Sections,
as we explicitly compute the pre-factors for the SO(N) case. 

\subsection{Second-order perturbation theory\label{subsec:Second-order-perturbation-theory-gen}}

We return to the 4-site chain, now with the assumption that the most
strongly coupled sites 2 and 3 have a singlet ground state. The singlet
ground state is trivial, in the sense of having no dynamics, and no
effective spin remains. This situation causes the first order perturbation
theory to vanish, as described above, and we have to rework the effective
problem to second-order in perturbation theory. The 4-site problem
becomes a two-site problem with site 1 effectively coupled to 4 (see
Fig.~\ref{fig:Schematic-decimation}). 

We call the singlet state $\left|\mathrm{s}\right\rangle $ and we
call $\mathcal{H}_{\Upsilon,\Upsilon'}^{\left(2\right)}$ the effective
Hamiltonian connecting sites 1 and 4 coming from tensors of rank $\Upsilon$
and $\Upsilon'$. By standard second order perturbation theory, $\mathcal{H}_{\Upsilon,\Upsilon'}^{\left(2\right)}$
reads

\begin{equation}
\mathcal{H}_{\Upsilon,\Upsilon'}^{\left(2\right)}=2K_{1}^{\Upsilon}K_{3}^{\Upsilon'}\sum_{\upsilon,\upsilon'}T_{\upsilon}^{\Upsilon}\left(\mathbf{L}_{1}\right)\Delta\mathcal{H}_{\upsilon,\upsilon'}^{\Upsilon\Upsilon'}T_{\upsilon'}^{\Upsilon'}\left(\mathbf{L}_{4}\right)\label{eq:second-order-start}
\end{equation}
where
\begin{equation}
\Delta\mathcal{H}_{\upsilon,\upsilon'}^{\Upsilon\Upsilon'}=\sum_{\tilde{\Upsilon},\tilde{\upsilon}}\frac{\left\langle \mathrm{s}\left|T_{\upsilon}^{\Upsilon}\left(\mathbf{L}_{2}\right)\right|\tilde{\Upsilon}\tilde{\upsilon}\right\rangle \left\langle \tilde{\Upsilon}\tilde{\upsilon}\left|T_{\upsilon'}^{\Upsilon'}\left(\mathbf{L}_{3}\right)\right|\mathrm{s}\right\rangle }{\Delta E_{\tilde{\Upsilon}}\left(\mathbf{L}_{2},\mathbf{L}_{3}\right)}.\label{eq:second-order-partial}
\end{equation}
The sum over $\tilde{\Upsilon}$ is over all representations arising
from the Clebsch-Gordan series of $\mathbf{L}_{2}\otimes\mathbf{L}_{3}$,
excluding the singlet. The energy denominator $\Delta E_{\tilde{\Upsilon}}$
is the difference between the energies of the singlet and that of
the multiplet $\tilde{\Upsilon}$. 

The main goal is to simplify Eq.~(\ref{eq:second-order-partial})
using all the selection rules available. Again, from Eq.~(\ref{eq:WE-theorem}),

\begin{align}
 & \left\langle \mathrm{s}\left|T_{\upsilon}^{\Upsilon}\left(\mathbf{L}_{2}\right)\right|\tilde{\Upsilon}\tilde{\upsilon}\right\rangle \nonumber \\
 & =\left\langle \Upsilon\tilde{\Upsilon},\mathrm{s}|\Upsilon\upsilon\tilde{\Upsilon}\tilde{\upsilon}\right\rangle \left\langle \Upsilon\left\Vert T_{\upsilon}^{\Upsilon}\left(\mathbf{L}_{2}\right)\right\Vert \tilde{\Upsilon}\right\rangle .
\end{align}

The coefficient $\left\langle \Upsilon\tilde{\Upsilon},\mathrm{s}|\Upsilon\upsilon\tilde{\Upsilon}\tilde{\upsilon}\right\rangle $
is non-vanishing only if the representations $\Upsilon$ and $\tilde{\Upsilon}$
have a singlet in their Clebsch-Gordan series. In this case, they
are called mutually complementary. For every Lie algebra, given a
representation $\Upsilon$, only one other unique representation $\bar{\Upsilon}$
exists that is complementary to it~\cite{Iachello_book}. For so(N)
and sp(N) algebras, every representation is complementary to itself,
$\bar{\Upsilon}=\Upsilon$. For the other cases, the complementary
to a given representation, though not necessarily equal to it, has
the same dimension. In su(N), for instance, the fundamental and anti-fundamental
representations generate singlets when combined. Therefore, the sum
over $\tilde{\Upsilon}$ is reduced to the single complementary representation
$\bar{\Upsilon}$. Applying this analysis to the matrix element of
the tensor living on site 3, $T_{\upsilon'}^{\Upsilon'}\left(\mathbf{L}_{3}\right)$,
we arrive at the selection rule $\Upsilon'=\bar{\Upsilon}$. Thus,

\begin{eqnarray}
 &  & \Delta\mathcal{H}_{\upsilon,\upsilon'}^{\Upsilon\Upsilon'}\nonumber \\
 & = & \delta_{\Upsilon',\bar{\Upsilon}}\sum_{\tilde{\upsilon}}\frac{\left\langle \mathrm{s}\left|T_{\upsilon}^{\Upsilon}\left(\mathbf{L}_{2}\right)\right|\bar{\Upsilon}\tilde{\upsilon}\right\rangle \left\langle \bar{\Upsilon}\tilde{\upsilon}\left|T_{\upsilon'}^{\bar{\Upsilon}}\left(\mathbf{L}_{3}\right)\right|\mathrm{s}\right\rangle }{\Delta E_{\bar{\Upsilon}}\left(\mathbf{L}_{2},\mathbf{L}_{3}\right)}\nonumber \\
 & = & \delta_{\Upsilon',\Upsilon}\frac{\sum_{\tilde{\upsilon}}\left\langle \mathrm{s}\left|T_{\upsilon}^{\Upsilon}\left(\mathbf{L}_{2}\right)\right|\Upsilon\tilde{\upsilon}\right\rangle \left\langle \Upsilon\tilde{\upsilon}\left|T_{\upsilon'}^{\Upsilon}\left(\mathbf{L}_{3}\right)\right|\mathrm{s}\right\rangle }{\Delta E_{\Upsilon}\left(\mathbf{L}_{2},\mathbf{L}_{3}\right)},
\end{eqnarray}
where, in the second equality, we have used the identity $\sum_{\tilde{\Upsilon},\tilde{\upsilon}}\left|\tilde{\Upsilon}\tilde{\upsilon}\right\rangle \left\langle \tilde{\Upsilon}\tilde{\upsilon}\right|=\mathbf{1}$
and finally particularized the result to SO(N) and Sp(N) by using
$\bar{\Upsilon}=\Upsilon$.

Using the decomposition of the identity operator 

\begin{equation}
\sum_{\tilde{\upsilon}}\left|\Upsilon\tilde{\upsilon}\right\rangle \left\langle \Upsilon\tilde{\upsilon}\right|=\mathbf{1}-\sum_{\tilde{\Upsilon}\ne\Upsilon,\tilde{\upsilon}}\left|\tilde{\Upsilon}\tilde{\upsilon}\right\rangle \left\langle \tilde{\Upsilon}\tilde{\upsilon}\right|,
\end{equation}
and using $\left\langle \tilde{\Upsilon}\tilde{\upsilon}\left|T_{\upsilon}^{\Upsilon}\left(\mathbf{L}_{2,3}\right)\right|\mathrm{s}\right\rangle =0$,
for $\tilde{\Upsilon}\ne\Upsilon$, we obtain
\begin{equation}
\Delta\mathcal{H}_{\upsilon,\upsilon'}^{\Upsilon\Upsilon'}=\frac{\delta_{\Upsilon',\Upsilon}}{\Delta E_{\Upsilon}\left(\mathbf{L}_{2},\mathbf{L}_{3}\right)}\left\langle \mathrm{s}\left|T_{2,\upsilon}^{\Upsilon}\left(\mathbf{L}_{2}\right)T_{3,\upsilon'}^{\Upsilon}\left(\mathbf{L}_{3}\right)\right|\mathrm{s}\right\rangle .\label{eq:second-order-interm}
\end{equation}
Now, from the symmetry properties of the Hamiltonian, preserved by
the SDRG, the effective Hamiltonian must read

\begin{equation}
\tilde{\mathcal{H}}\propto\mathcal{O}^{\Upsilon}\left(\mathbf{L}_{1},\mathbf{L}_{4}\right),
\end{equation}
since this is the only symmetric scalar operator that can be built
out of $\Upsilon$-rank tensors. The remaining matrix element can
thus be computed to give~\cite{Iachello_book}

\begin{align}
\left\langle \mathrm{s}\left|T_{2,\upsilon}^{\Upsilon}\left(\mathbf{L}_{2}\right)T_{3,\upsilon'}^{\Upsilon}\left(\mathbf{L}_{3}\right)\right|\mathrm{s}\right\rangle  & =\delta_{\upsilon',-\upsilon}\left(-1\right)^{f\left(\Upsilon,\upsilon\right)}\nonumber \\
 & \times\alpha\left(\Upsilon,\mathbf{L}_{2},\mathbf{L}_{3}\right),
\end{align}
where $\alpha\left(\Upsilon,\mathbf{L}_{2},\mathbf{L}_{3}\right)$,
the reduced matrix element, is a function of the tensor rank $\Upsilon$
and the spins being decimated. Its explicit value will be determined
for the cases of interest. Collecting the results and plugging them
back into Eq.~(\ref{eq:second-order-start}), we arrive at the effective
Hamiltonian connecting sites 1 and 4

\begin{eqnarray}
\mathcal{H}_{\Upsilon,\Upsilon'}^{\left(2\right)} & \equiv & \frac{2\delta_{\Upsilon',\Upsilon}\alpha\left(\Upsilon,\mathbf{L}_{2},\mathbf{L}_{3}\right)}{\Delta E_{\Upsilon}\left(\mathbf{L}_{2},\mathbf{L}_{3}\right)}K_{1}^{\Upsilon}K_{3}^{\Upsilon}\times\nonumber \\
 & \times & \sum_{\upsilon}T_{\upsilon}^{\Upsilon}\left(\mathbf{L}_{1}\right)\left(-1\right)^{f\left(\Upsilon,\upsilon\right)}T_{-\upsilon}^{\Upsilon}\left(\mathbf{L}_{4}\right)\nonumber \\
 & = & \frac{2\delta_{\Upsilon',\Upsilon}\alpha\left(\Upsilon,\mathbf{L}_{2},\mathbf{L}_{3}\right)}{\Delta E_{\Upsilon}\left(\mathbf{L}_{2},\mathbf{L}_{3}\right)}K_{1}^{\Upsilon}K_{3}^{\Upsilon}\mathcal{O}^{\Upsilon}\left(\mathbf{L}_{1},\mathbf{L}_{4}\right),\label{eq:K-order2-a}
\end{eqnarray}
where in the last step we have used Eq.~(\ref{eq:CG_tensors}) to
identify $\mathcal{O}^{\Upsilon}\left(\mathbf{L}_{1},\mathbf{L}_{4}\right)$. 

This derivation guarantees that tensors of different ranks again \emph{do
not get mixed by the SDRG}, which simplifies the analysis of the flow
dramatically. In fact, this is the advantage of working with irreducible
tensors \cite{quitoprb2016}.\textcolor{red}{{} }Also, the functional
form of the Hamiltonian does not change by the decimation steps. This
is schematically represented in Fig.~\ref{fig:Schematic-decimation}.
Summing up, the coupling constants renormalize according to

\begin{equation}
\tilde{K}_{1,4}^{\Upsilon}=\frac{2\alpha\left(\Upsilon,\mathbf{L}_{2},\mathbf{L}_{3}\right)}{\Delta E_{\Upsilon}\left(\mathbf{L}_{2},\mathbf{L}_{3}\right)}K_{1}^{\Upsilon}K_{3}^{\Upsilon}.\label{eq:K-order2}
\end{equation}
This is the generalization to any symmetry group of the SDRG step
first derived for SU(2)-symmetric spin-1/2 chains in Refs.~\onlinecite{madasgupta,madasguptahu}
and generalized to any SU(2) spin in Ref.~\onlinecite{westerbergetal}.

The application of the SDRG has thus been generalized for spin chains
of any Lie group symmetry. The main ingredients for this are the identification
of the tensor operator technology, and the Wigner-Eckart theorem.
In what follows, we apply the formulas to the SO$\left(N\right)$
and Sp$\left(N\right)$ cases of our interest, finding closed expressions
for the pre-factors .

\subsection{SO(N) rules in closed form \label{sec:RG-Steps-closed-form}}

Let us particularize Eqs.~(\ref{eq:K-order-1}) and (\ref{eq:K-order2}),
which dictate how the couplings are renormalized in SDRG steps, to
the case of SO(N)-symmetric Hamiltonians. To do so, we start with
Eq.~(\ref{eq:HamiltSON}) for the strongly coupled sites 2 and 3
\begin{equation}
\mathcal{H}_{2,3}=K_{2}^{\left(1\right)}\mathcal{O}^{\left(1\right)}\left(\mathbf{L}_{2},\mathbf{L}_{3}\right)+K_{2}^{\left(2\right)}\mathcal{O}^{\left(2\right)}\left(\mathbf{L}_{2},\mathbf{L}_{3}\right).\label{eq:SON-strongly-coupled-pair}
\end{equation}
In what follows, it proves useful to rewrite the two coupling constants
of each bond $\left(K_{i}^{\left(1\right)},K_{i}^{\left(2\right)}\right)$
in polar coordinates. In particular, as we will see, the ratio of
Eqs.~(\ref{eq:K-order-1}) and (\ref{eq:K-order2}) fully controls
the renormalization of the angles $\theta_{i}$

\begin{equation}
\tan\theta_{i}=\frac{K_{i}^{\left(2\right)}}{K_{i}^{\left(1\right)}},\label{eq:angle-variables}
\end{equation}
while the radial variable 

\begin{equation}
r_{i}=\sqrt{\left(K_{i}^{\left(1\right)}\right)^{2}+\left(K_{i}^{\left(2\right)}\right)^{2}}\label{eq:radial-variable}
\end{equation}
controls the energy scale.

It will be important for our later discussion to know that some points
of the parameter space have, in fact, SU(N) symmetry. First, $K_{i}^{\left(1\right)}=K_{i}^{\left(2\right)}$
is an obvious SU(N)-symmetric point, where the Hamiltonian becomes

\begin{equation}
\mathcal{H}_{2,3}=K_{2}^{\left(1\right)}\sum_{\mu=1}^{N^{2}-1}\Lambda_{2}^{\mu}\Lambda_{3}^{\mu},
\end{equation}
which is the Heisenberg SU(N) Hamiltonian at sites $\left(2,3\right)$.
This corresponds to the $\theta=\frac{\pi}{4}$ point in the polar
coordinates of Eq.~(\ref{eq:angle-variables}). 

The choice $K_{i}^{\left(1\right)}=-K_{i}^{\left(2\right)}$ is also
SU(N) symmetric ($\theta=-\frac{\pi}{4}$). This can be shown in the
following way. Starting from the SU$\left(N\right)$ invariant point
$\theta=\frac{\pi}{4}$, we transform all SU(N) generators as $\Lambda_{i}^{a}\rightarrow-\Lambda_{i}^{a*}$
on every other site. This changes the corresponding SU(N) representation
from the fundamental to the anti-fundamental, which is its complex-conjugate.
To show the SU(N) invariance, recall that $\mathcal{O}^{\left(1\right)}$
is built with the generators of SO(N), which are purely imaginary
antisymmetric objects and, therefore, do not change sign under this
transformation. Meanwhile, all the terms in $\mathcal{O}^{\left(2\right)}$
are constructed using the real generators of SU(N) and will, therefore,
flip sign. By absorbing this sign change into $K_{i}^{\left(2\right)}$,
we see that the point $\theta=-\frac{\pi}{4}$ is also SU(N) symmetric.

Notice that the derivation of last Section guarantees that no operators
other than $\mathcal{O}^{\left(1\right)}$ and $\mathcal{O}^{\left(2\right)}$
will be generated during the RG flow. This was not obvious were we
not aware that the SU(N) anisotropy keeps the underlying SO(N) structure
intact. We will, in what follows, make full use of the fact that the
SO(N)-symmetric Hamiltonian can be thought of as an anisotropic SU(N)
Hamiltonian and also that the renormalization pre-factors are determined
by very few quantities: (i) the representations $\Upsilon_{2}$ and
$\Upsilon_{3}$, of spins on sites $2$ and $3$ and (ii) the two-spin
ground manifold $\tilde{\Upsilon}$. 

\subsubsection{First-order perturbation theory}

According to Eqs.~(\ref{eq:K-order-1}), for both links 1 and 3,
the renormalization of $\tan\theta_{i}$ (the ratio of couplings)
is given by, 

\begin{equation}
\tan\tilde{\theta}_{1,3}=\frac{\beta_{1,3}\left(2,\tilde{\Upsilon},\Upsilon_{2},\Upsilon_{3}\right)}{\beta_{1,3}\left(1,\tilde{\Upsilon},\Upsilon_{2},\Upsilon_{3}\right)}\tan\theta_{1,3},
\end{equation}
To compute these we take a shortcut using the results of Section~\ref{sec:SO(N)-group}.
Since the SDRG preserves the SU(N) symmetry, we expect that if we
start with all angles equal to $\frac{\pi}{4}$, $\tilde{\theta}$
has to be equal to $\pm\frac{\pi}{4}$. From that, we find that the
only possible values for the ratios of $\tan\tilde{\theta}_{1,3}/\tan\theta_{1,3}$
are $\pm1$, with the sign depending on the representations. At this
stage, we parametrize 

\begin{eqnarray}
\tilde{K}_{1,3}^{\left(1\right)} & = & \Phi_{1,3}K_{1,3}^{\left(1\right)},\label{eq:firstorderdecimation1}\\
\tilde{K}_{1,3}^{\left(2\right)} & = & \Phi_{1,3}\xi_{1,3}K_{1,3}^{\left(2\right)},\label{eq:firstorderdecimation2}
\end{eqnarray}
with $\xi_{1,3}\left(\tilde{\Upsilon},\Upsilon_{2},\Upsilon_{3}\right)=\pm1$
and $\Phi_{1,3}\left(\tilde{\Upsilon},\Upsilon_{2},\Upsilon_{3}\right)$
to be determined according to the representations. 

In order to determine $\xi$ and $\Phi$, we identify the representations
$\Upsilon$ in a Young-tableau notation for SO(N). Since we focus
on phase on which only anti-symmetric representations are generated
by the SDRG flow, we specify them by $Q$, the number of vertically
concatenated boxes. We also chose for concreteness $Q_{2}\le Q_{3}$
and assumed that the two-spin ground state $\tilde{Q}$ is not a singlet,
so that first-order renormalization is required. All possible relevant
cases are listed in Table~\ref{tab:List-of-beta}, with the values
of $\xi$ and $\Phi$ provided. In general, sign flips always happen
on the bond on the side of the smaller $Q_{i}$ ($i=2,3$) of the
decimated pair. The derivation of the values of $\xi_{1},\,\xi_{3},\,\Phi_{1}$
and $\Phi_{3}$ is given in the Appendix~\ref{sec:Derivation-xi-Phi}.

\begin{widetext}
\begin{center}
\begin{table}
\begin{centering}
\begin{tabular}{|c|c|c|c|c|}
\hline 
 & $\xi_{1}$ & $\xi_{3}$ & $\Phi_{1}$ & $\Phi_{3}$\tabularnewline
\hline 
\hline 
$\tilde{Q}=Q_{2}+Q_{3}$ and $Q_{2}+Q_{3}\le\text{int}\left(\frac{\mathrm{N}}{2}\right)$  & $1$ & $1$ & $\frac{Q_{2}}{Q_{2}+Q_{3}}$ & $\frac{Q_{3}}{Q_{2}+Q_{3}}$\tabularnewline
\hline 
$\tilde{Q}=\mathrm{N}-\left(Q_{2}+Q_{3}\right)$ and $Q_{2}+Q_{3}>\text{int}\left(\frac{\mathrm{N}}{2}\right)$  & $-1$ & $-1$ & $\frac{Q_{2}}{Q_{2}+Q_{3}}$ & $\frac{Q_{3}}{Q_{2}+Q_{3}}$\tabularnewline
\hline 
$\tilde{Q}=\left|Q_{3}-Q_{2}\right|$  & $-1$ & $1$ & $\frac{Q_{2}}{\mathrm{N}-Q_{2}+Q_{3}}$ & $\frac{\mathrm{N}-Q_{3}}{\mathrm{N}-Q_{2}+Q_{3}}$\tabularnewline
\hline 
\end{tabular}
\par\end{centering}
\caption{List of the pre-factors $\xi_{1},$ $\xi_{3},$$\Phi_{1}$ and $\Phi_{3}$
used in the first-order decimations. \label{tab:List-of-beta}}
\end{table}
\par\end{center}

\end{widetext}

\subsubsection{Second-order perturbation theory \label{subsec:Second-order-perturbation-theory}}

Particularizing the second-order SDRG decimation derived in Section~\ref{subsec:Second-order-perturbation-theory-gen}
to SO(N), the effective Hamiltonian between sites 1 and 4 acquires
the following form

\begin{eqnarray}
\Delta\mathcal{H}_{\left(1,4\right)}^{\left(2\right)} & = & \tilde{K}_{1,4}^{\left(1\right)}\mathcal{O}^{\left(1\right)}\left(\mathbf{L}_{1},\mathbf{L}_{4}\right)+\tilde{K}_{1,4}^{\left(2\right)}\mathcal{O}^{\left(2\right)}\left(\mathbf{L}_{1},\mathbf{L}_{4}\right).
\end{eqnarray}
Going back to Eq.~(\ref{eq:K-order2}), we can write explicitly
\begin{eqnarray}
\tilde{K}_{1,4}^{\left(1\right)} & = & \alpha^{\left(1\right)}\left(\mathbf{L}_{2},\mathbf{L}_{3}\right)\frac{K_{1}^{\left(1\right)}K_{3}^{\left(1\right)}}{E_{\left[0,0\right]}-E_{\left[1,1\right]}},\\
\tilde{K}_{1,4}^{\left(2\right)} & = & \alpha^{\left(2\right)}\left(\mathbf{L}_{2},\mathbf{L}_{3}\right)\frac{K_{1}^{\left(2\right)}K_{3}^{\left(2\right)}}{E_{\left[0,0\right]}-E_{\left[2,0\right]}},
\end{eqnarray}
with $\alpha^{\left(1,2\right)}$ yet to be determined. The energies
in the denominators come from the spectrum of the strongly coupled
pair of sites $\left(2,3\right)$, Eq.~(\ref{eq:SON-strongly-coupled-pair}).
$E_{\left[0,0\right]}$ is the energy of the singlet representation
$\left[0,0\right]$, while the energies $E_{\left[1,1\right]}$ and
$E_{\left[2,0\right]}$ correspond to the representations complementary
to the ones of the operators forming the scalars $\mathcal{O}^{\left(1\right)}$
and $\mathcal{O}^{\left(2\right)}$, respectively. As pointed out
before, for SO(N), these are just the same as the ranks of the scalars
themselves.

A key observation is that the gaps $E_{\left[0,0\right]}-E_{\left[1,1\right]}$
and $E_{\left[0,0\right]}-E_{\left[2,0\right]}$ close at the point
where the $\left[0,0\right]$ and $\left[1,1\right]$ representations
cross (a generalization of the AKLT point of the SU(2) spin-1 chain~\cite{PhysRevLett.59.799})
and at the SU(N)-invariant points, respectively. Since the energy
denominators are linear in $K_{2}^{\left(1\right)}$ and $K_{2}^{\left(2\right)}$
, we must have

\begin{align}
E_{\left[0,0\right]}-E_{\left[1,1\right]} & \propto K_{2}^{\left(1\right)}\left(\tan\theta_{AKLT}-\tan\theta_{2}\right),\\
E_{\left[0,0\right]}-E_{\left[2,0\right]} & \propto K_{2}^{\left(1\right)}\left(\tan\theta_{SU(N)}-\tan\theta_{2}\right).
\end{align}
 The generalized AKLT point is known for SO(N) systems to be given
by~\cite{Tu_PhysRevB.78.094404}

\begin{equation}
\tan\theta_{AKLT}=\frac{\mathrm{N}-2}{\mathrm{N}+2},\label{eq:AKLTangle}
\end{equation}
while the SU(N)-symmetric point where the representations $\left[0,0\right]$
and $\left[2,0\right]$ meet in energy is antipodal to $\pi/4$,

\begin{equation}
\theta_{SU(N)}=-\frac{3\pi}{4}.
\end{equation}

At this stage, the rules are simplified to

\begin{eqnarray}
\tilde{K}_{1,4}^{\left(1\right)} & = & \tilde{\alpha}^{\left(1\right)}\left(\mathbf{L}_{2},\mathbf{L}_{3}\right)\frac{K_{1}^{\left(1\right)}K_{3}^{\left(1\right)}}{K_{2}^{\left(1\right)}\left(\tan\theta_{AKLT}-\tan\theta_{2}\right)},\label{eq:order-2-RG-rule-p1}\\
\tilde{K}_{1,4}^{\left(2\right)} & = & \tilde{\alpha}^{\left(2\right)}\left(\mathbf{L}_{2},\mathbf{L}_{3}\right)\frac{K_{1}^{\left(2\right)}K_{3}^{\left(2\right)}}{K_{2}^{\left(1\right)}\left(1-\tan\theta_{2}\right)},\label{eq:order-2-RG-rule-p2}
\end{eqnarray}
with the remaining task of determining the newly defined $\tilde{\alpha}^{\left(1,2\right)}$.
For that, we can take advantage again of the presence of an SU(N)-symmetric
point in the phase diagram. At the SU(N)-symmetric point $-\frac{\pi}{4}$,
$K_{i}^{\left(1\right)}=-K_{i}^{\left(2\right)}$ and, once again
enforcing that the SDRG must preserve the SU(N) symmetry, $\tilde{K}_{1,4}^{\left(1\right)}=-\tilde{K}_{1,4}^{\left(2\right)}$.
Dividing Eqs.~(\ref{eq:order-2-RG-rule-p1}) and (\ref{eq:order-2-RG-rule-p2}),
we conclude that 

\begin{equation}
\frac{\tilde{\alpha}^{\left(1\right)}\left(\mathbf{L}_{2},\mathbf{L}_{3}\right)}{\tilde{\alpha}^{\left(2\right)}\left(\mathbf{L}_{2},\mathbf{L}_{3}\right)}=-\frac{\left(\tan\theta_{AKLT}+1\right)}{2}.\label{eq:second-order-ratio}
\end{equation}
Once the ratio is fixed, the value of $\tilde{\alpha}^{\left(2\right)}$
can be found by comparing Eq.~(\ref{eq:order-2-RG-rule-p2}) with
the RG step for SU(N)-symmetric chains from reference Ref.~\onlinecite{PhysRevB.70.180401},

\begin{equation}
\tilde{\alpha}^{\left(2\right)}\left(\mathbf{L}_{2},\mathbf{L}_{3}\right)=-\frac{4Q\left(\mathrm{N}-Q\right)}{\mathrm{N}^{2}\left(\mathrm{N}-1\right)},
\end{equation}
where $Q_{2}=Q_{3}=Q$ is the number of boxes in the Young tableaux
of SO(N) at sites 2 and 3. Recall that a necessary condition for singlet
formation is that the same representation appear on both sites. Putting
everything together, we get

\begin{eqnarray}
\tilde{K}_{1,4}^{\left(1\right)} & = & \left[\frac{4Q\left(\mathrm{N}-Q\right)}{\left(\mathrm{N}-1\right)\mathrm{N}\left(\mathrm{N}+2\right)}\right]\frac{K_{1}^{\left(1\right)}K_{3}^{\left(1\right)}}{K_{2}^{\left(1\right)}\left(\frac{\mathrm{N}-2}{\mathrm{N}+2}-\tan\theta_{2}\right)},\label{eq:K-order2-1}\\
\tilde{K}_{1,4}^{\left(2\right)} & = & -\left[\frac{4Q\left(\mathrm{N}-Q\right)}{\mathrm{N}^{2}\left(\mathrm{N}-1\right)}\right]\frac{K_{1}^{\left(2\right)}K_{3}^{\left(2\right)}}{K_{2}^{\left(1\right)}\left(1-\tan\theta_{2}\right)}.\label{eq:K-order2-2}
\end{eqnarray}

The renormalization of the angle is found by dividing Eq.~(\ref{eq:K-order2-1})
by (\ref{eq:K-order2-2}) 

\begin{eqnarray}
\tan\tilde{\theta}_{1,4} & = & -\left(\frac{\mathrm{N}+2}{\mathrm{N}}\right)\frac{\frac{\mathrm{N}-2}{\mathrm{N}+2}-\tan\theta_{2}}{1-\tan\theta_{2}}\tan\theta_{1}\tan\theta_{3}.\label{eq:angle-RG-2-order}
\end{eqnarray}
One can verify explicitly from Eq.~(\ref{eq:angle-RG-2-order}) the
existence of \emph{angular fixed points}. These points are such that 

\begin{equation}
\tan\tilde{\theta}_{1,4}=\tan\theta_{i},\,i=1,2,3.
\end{equation}
Besides the SU(N)-symmetric point $\tan\theta_{i}=-1$ $\left(\theta_{i}=-\frac{\pi}{4}\right)$,
by using Eq.~(\ref{eq:angle-RG-2-order}), we find that $\tan\theta_{i}=0$
$\left(\theta_{i}=0\right)$ and $\tan\theta_{i}\rightarrow-\infty$
$\left(\theta_{i}=-\frac{\pi}{2}\right)$ are also angular fixed points.
At this stage, we are also able to determine their stability. By including
a perturbation $\delta\theta_{i}$ to the fixed points and by expanding
Eq.~(\ref{eq:angle-RG-2-order}) in powers of $\delta\theta$, we
find that $\theta_{i}=0$ and $\theta_{i}=-\frac{\pi}{2}$ are stable,
whereas $\theta_{i}=-\frac{\pi}{4}$ is unstable. This is expected
since they are SU(N)-symmetric points and deviations from this symmetry
are expected to be amplified by the SDRG. Notice that in this analysis
we assume that only second-order decimation occurs, which can be achieved
by the choice of the initial angle distribution, as we will show later.
In the case where Eq.~(\ref{eq:angle-RG-2-order}) leads also to
first order decimations, the representations will also flow, and the
analysis of the angular fixed points as well as their stability is
more elaborate. We postpone this analysis for later.

\subsection{Sp(N) rules in closed form \label{sec:RG-Steps-closed-form-Sp(N)}}

The derivation of the Sp(N) rules are analogous to the SO(N) ones
in as far as only second-order decimations are present. We can again
use the shortcut of having an SU(N)-symmetric point to explicitly
compute the necessary pre-factors. Just as in the SO(N) case, we work
with polar coordinates with the angle $\theta_{i}$ defined as $\tan\theta_{i}=\frac{K_{i}^{\left(2\right)}}{K_{i}^{\left(1\right)}}$.
These rules allow us to completely characterize the physics of a large
fraction of the Sp(N) AF phase diagram and read

\begin{eqnarray}
\tilde{K}_{1,4}^{\left(1\right)} & = & \left(\frac{4Q^{2}}{\left(\mathrm{N}-1\right)\mathrm{N}\left(\mathrm{N}+2\right)}\right)\frac{K_{1}^{\left(1\right)}K_{3}^{\left(1\right)}}{K_{2}^{\left(1\right)}\left(\frac{\mathrm{N}-2}{\mathrm{N}+2}-\tan\theta_{2}\right)},\\
\tilde{K}_{1,4}^{\left(2\right)} & = & -\left(\frac{4Q^{2}}{\mathrm{N}^{2}\left(\mathrm{N}-1\right)}\right)\frac{K_{1}^{\left(1\right)}K_{3}^{\left(1\right)}}{K_{2}^{\left(1\right)}\left(1-\tan\theta_{2}\right)},
\end{eqnarray}
where $Q$ denotes the number of boxes in the Sp(N) Young tableaux
at sites 2 and 3.

The situation is different, however, when first order decimations
in the AF region are also required. The reason is that this region
has no SU(N)-symmetric point. That implies that the above shortcut
of using these points to compute the prefactors is no longer valid.
We can, however, build the Sp(N) tensors explicitly, and from that,
calculate all the necessary prefactors. We will come back to this
point when we study the Sp(N) SDRG flow.

\section{The phase diagram of strongly disordered SO(N) and Sp(N) spin chains\label{sec:subgroups-1-1-1}}

With the decimation rules of the previous Section, we are now able
to characterize the SDRG flow for SO(N) and Sp(N) chains. The characterization
of the RG flow involves finding the low-energy behavior of the joint
distribution of $\theta$, $r$ and the representations $\Upsilon$
at energy scale $\Omega$, $P\left(r,\theta,\Upsilon;\Omega\right)$.
At the beginning of the flow we set $\Omega=\Omega_{0}$ and assume
the initial distribution to be separable, $P\left(r,\theta,\Upsilon;\Omega_{0}\right)=P_{r}\left(r\right)P_{\theta}\left(\theta\right)P_{\Upsilon}\left(\Upsilon\right)$.
The initial distribution for the angles is chosen to be a delta function

\begin{equation}
P_{\theta}\left(\theta\right)=\delta\left(\theta-\theta_{0}\right).
\end{equation}
Similarly, the initial choice of representations is not random but
fixed at the defining representation, as mentioned previously,
\begin{equation}
P_{\Upsilon}\left(\Upsilon\right)=\delta\left(\Upsilon-\left[1,\vec{0}\right]\right).\label{eq:irrep_dist}
\end{equation}
Initial disorder is present in the radial variable $r$, through a
finite standard deviation for $P_{r}\left(r\right)$. We also assume
that the initial disorder distribution width is sufficiently large
for the SDRG method to be applicable.

Let us now list some universal features of the SDRG flow in the antiferromagnetic
(AF) phases.\footnote{We use the term antiferromagnetic here in the loose sense that there
is a tendency to form singlets. Actually, there is never true antiferromagnetic
order} During the initial stages of the flow, the distributions of angles,
radii and representations become correlated, but eventually become
again uncorrelated at low energies $\Omega\ll\Omega_{0}$. Furthermore,
at low energies, all angles tend to a single value, that is, $P_{\theta}$
flows to a delta function centered at one of the angular fixed points.
This is the main advantage of parametrizing the SDRG flow using polar
coordinates.~\cite{Quito_PhysRevLett.115.167201} As for the radial
distribution, it will flow at low energies to an infinite-disorder
profile, that is, its standard deviation divided by its average diverges.
More specifically,\cite{madasguptahu,madasgupta,fisher94-xxz}
\begin{equation}
P_{r}\left(r\right)\to\frac{1}{r^{1-\alpha}},\,\,\ \ \alpha^{-1}\sim\ln\left(\frac{\Omega_{0}}{\Omega}\right).
\end{equation}

In the phases discussed in this work, only anti-symmetric representations
are generated and the number of such distinct representations is finite.
The frequency of distinct representations at low energies depends
on $N$ and on the region of the parameter space.

At low energies, the remaining non-decimated spins are embedded in
a ``soup'' of randomly located singlets with a wide distribution of
sizes. At energy scale $\Omega$, the average separation of non-decimated
spins scales as $L_{\Omega}\sim\left|\ln\Omega\right|^{1/\psi}$,
with the exponent $\psi$ depending on the number of distinct representations
at the low-energy fixed point.~\cite{Damle2002,PhysRevB.70.180401}
At a scale $\Omega\sim T$, only non-decimated spins contribute to
the susceptibility. Since the distributions of couplings is extremely
broad, the spins will be typically very weakly coupled $r\ll T$.
Thus, the susceptibility is given by Curie's law: $\chi\left(T\right)^{-1}\sim TL_{T}\sim T\left|\ln T\right|^{1/\psi}$.\textcolor{black}{\cite{fisher94-xxz,QuitoLopes2017PRL}}
Other thermodynamic quantities follow from similar reasoning: \textcolor{black}{the
entropy density is $s\left(T\right)\sim\left(\ln\mathrm{N}\right)/L_{T}$
and the specific heat $c\left(T\right)=T\left(ds/dT\right)\sim\left|\ln T\right|^{-1-1/\psi}$.}

The ground state consists of singlets formed with \emph{all} representations.
These singlets are formed at various stages of the SDRG. On average,
their sizes and strengths reflect the stage at which they were formed.
Therefore, singlets coupled with strength $r$ (i.e., whose higher
multiplets are excited at this energy cost) have sizes $L_{r}\sim\left|\ln r\right|^{1/\psi}$.
This picture motivates the name ``random singlet phase'' (RSP).\cite{fisher94-xxz,fishertransising2,PhysRevB.70.180401}

In what follows, we add more detail to the above general picture and
determine the phase diagrams of SO(N)-symmetric disordered chains.
We do so separately for the cases of odd and even N as their analyses
are different. Moreover, a more illuminating route consists of considering
specific small values of N first and then generalizing to the larger
ones. SO(2)-invariant Hamiltonians, obtained by including anisotropy
terms in SU(2) Hamiltonians (i. e. the XXZ model), were first studied
by Fisher.~\cite{fisher94-xxz} Generic SO(3) Hamiltonians were studied
more recently,~\cite{Quito_PhysRevLett.115.167201} in the context
of spin-1 SU(2) invariant Hamiltonians. Thus, we use the SDRG flows
of the groups SO(4) and SO(5) as the simplest yet unexplored examples
of orthogonal symmetric Hamiltonians. Then, the extrapolation to arbitrary
N is found to be straightforward. The $\mathrm{N}=4$ case presents
a feature that is not present for any other N, which makes the construction
of its phase diagram slightly more subtle. For this reason, we will
take up the odd N case first. For symplectic invariant Hamiltonians,
we are not aware of any previous analysis via the SDRG. We focus on
discussions of Sp(4) and Sp(6), and provide a brief analysis and expectations
for larger N.

\subsection{SO(N) for odd values of N}
\begin{center}
\begin{figure}
\begin{centering}
\includegraphics[width=0.85\columnwidth]{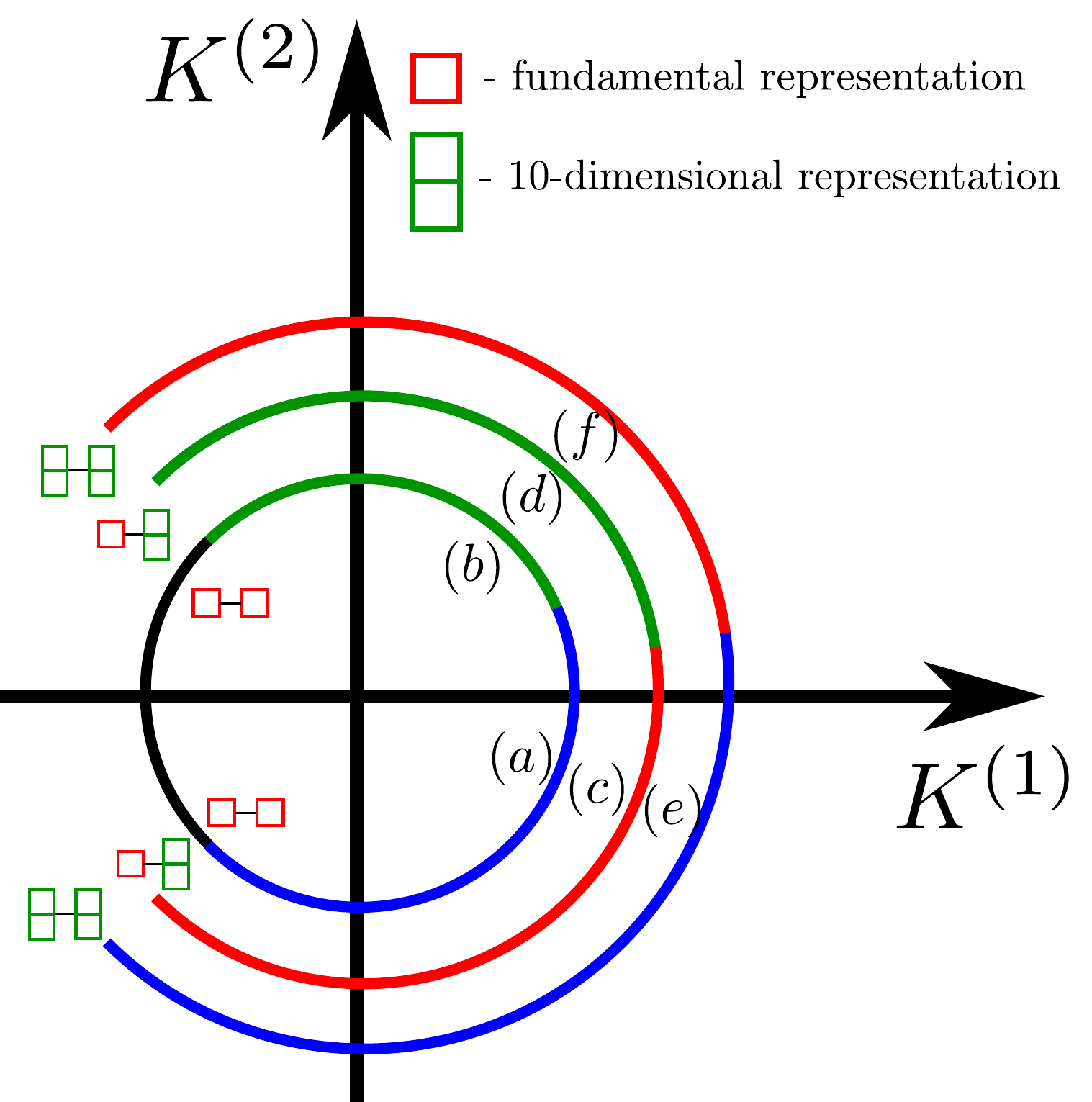}
\par\end{centering}
\caption{(Color online) Ground state structure of the SO(5) two-site problem
as a function of the angle $\theta$, with different colors representing
the distinct multiplets. Each arc of fixed radius corresponds to the
ground state multiplets for the two representations indicated next
to it by Young tableaux. The singlet state is represented in blue,
the fundamental 5-dimensional $\left[1,0\right]$ multiplet in red
{[}represented by a single box, in SO(N) Young-tableau notation{]}
and the 10-dimensional $\left[1,1\right]$ multiplet in green {[}represented
by two boxes concatenated vertically{]}. All other multiplets are
colored black, and do not participate in the flow of anti-ferromagnetic
phases. The points with $K^{\left(1\right)}>0$ where there is a change
in the ground state are generally called $\theta_{AKLT}$. The other
points where there is a change in the ground state for $K^{\left(1\right)}<0$
are $\pm3\pi/4$ (for any pair of SO(5) representations displayed).
The RG rules corresponding to the letters $(a)-(f)$ are given in
Fig.~\ref{fig:All-decimations-SO(5)} \label{fig:levels-so5}}
\end{figure}
\par\end{center}

To begin the analysis of odd-N SO(N)-symmetric disordered chains,
we focus first on SO(5). In Fig.~\ref{fig:levels-so5}, we represent
the ground states of the two-spin problems as functions of $\theta$
using a color code. The important representations and their respective
colors for the AF SDRG flow are the singlet $\left[0,0\right]$ (blue),
the fundamental $\left[1,0\right]$ (red) and $\left[1,1\right]$
(green). As discussed, the non-trivial ones have dimensions $d_{\left[1,0\right]}=5$
and $d_{\left[1,1\right]}=10$ and, in a Young-tableau language, are
represented by a single box and by two vertically stacked boxes, respectively.
Each circle represents a particular combination of representations
for the two spins that are coupled, as indicated by the pair of Young
tableaux next to it. The colors of the arcs of each circle indicate,
through the color code, to which representation the ground multiplet
belongs. 

\begin{widetext}

\begin{figure}
\centering{}\includegraphics[width=1\columnwidth]{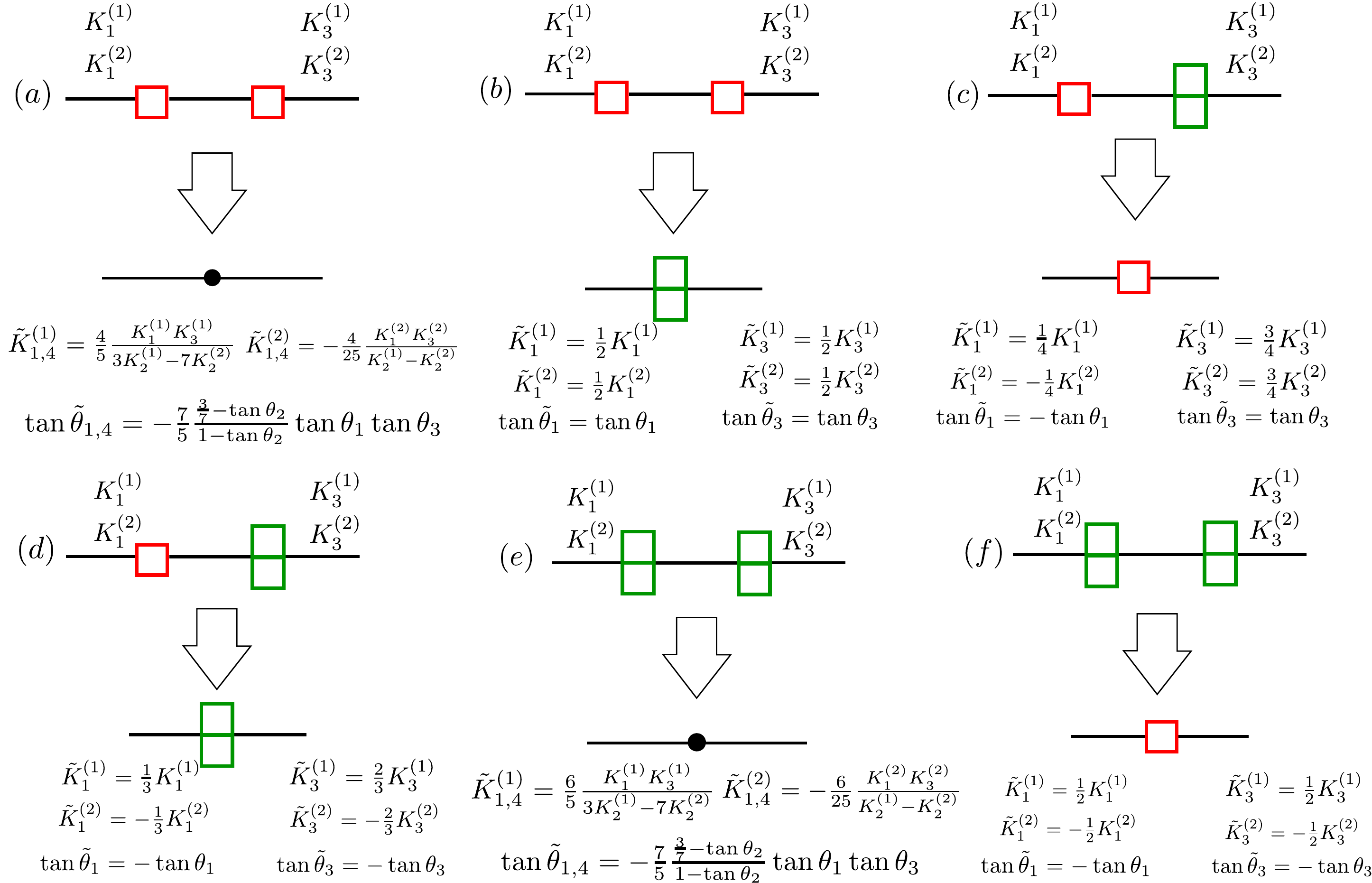}\caption{All possible decimations and RG rules needed for the antiferromagnetic
phases of the SO(5) model, representing using the Young-tableau notation.
\label{fig:All-decimations-SO(5)}}
\end{figure}

\end{widetext}

The innermost circle in Fig.~\ref{fig:levels-so5} displays the possible
ground states when the two spins are in the fundamental (defining)
representation. The points where colors change, and so does the two-spin
problem ground multiplet, are $-3\pi/4$, $3\pi/4$, for $K^{\left(1\right)}<0$,
and the angle $\theta_{AKLT}=\arctan\left(3/7\right)$ {[}see Eq.~(\ref{eq:AKLTangle}){]},
for $K^{\left(1\right)}>0$. We see that there are three possible
representations for the ground multiplet: the totally symmetric, 14-dimensional
$\left[2,0\right]$ (black), the $\left[0,0\right]$ singlet (blue),
and the $\left[1,1\right]$ (green). Since in this work we do not
focus on symmetric representations (analogous to the formation of
``large spins'' in the SU(2) case), we will neglect decimations in
the black region. If a decimation in the chain is performed in the
blue region, the two fundamental representation spins are substituted
by a singlet and the neighboring couplings are renormalized in second
order of perturbation theory. If a decimation is made in the green
region, on the other hand, the two $\left[1,0\right]$ spins are substituted
by a single spin in the 10-dimensional $\left[1,1\right]$ representation
and the renormalization of couplings is given by first-order perturbation
theory. This representation, at a later point of the RG flow, will
be decimated either with another fundamental $\left[1,0\right]$ object,
or with another $\left[1,1\right]$ object. This is why we need the
other circles in Fig.~\ref{fig:levels-so5}. For SO(5), the relevant
Clebsch-Gordan series are, therefore,~\cite{FegerTables}

\begin{eqnarray}
\left[1,0\right]\otimes\left[1,0\right] & = & \boldsymbol{\left[0,0\right]}\oplus\boldsymbol{\left[1,1\right]}\oplus\left[2,0\right]\label{eq:CG_1}\\
\left[1,0\right]\otimes\left[1,1\right] & = & \boldsymbol{\left[1,0\right]}\oplus\boldsymbol{\left[1,1\right]}\oplus\left[2,1\right]\label{eq:CG_2}\\
\left[1,1\right]\otimes\left[1,1\right] & = & \boldsymbol{\left[0,0\right]}\oplus\boldsymbol{\left[1,0\right]}\oplus\left[1,1\right]\oplus\nonumber \\
 & \oplus & \left[2,0\right]\oplus\left[2,1\right]\oplus\left[2,1\right].\label{eq:CG_3}
\end{eqnarray}
The bold terms on the right-hand side are the relevant ones for the
AF SDRG flow. For concreteness, we give in Fig.~\ref{fig:All-decimations-SO(5)}
all the RG rules for AF decimations in SO(5). By combining Fig.~\ref{fig:levels-so5}
and Fig.~\ref{fig:All-decimations-SO(5)}, we can characterize the
SO(5) RG flow, as we explain next.

The initial condition of starting with the fundamental representation
of SO(5) implies that the initial two-site ground-state structure
relevant to us is the one in the innermost arc of Fig.~\ref{fig:levels-so5}.Two
possibilities follow next, depending on what the initial angle $\theta_{0}$
is. If the angle $\theta_{0}$ is restricted to the blue region of
Fig.~\ref{fig:levels-so5} ($-\frac{3}{4}\pi<\theta_{0}<\theta_{AKLT}$),
only singlets are formed throughout the flow {[}only decimation (a)
in Fig.~\ref{fig:All-decimations-SO(5)} happens{]}. This is because
the renormalized angles also lie within the same blue range, as one
can explicitly verify using the angular equation in Fig.~\ref{fig:All-decimations-SO(5)}(a).
The distribution of angles has, therefore, to flow to one of the possible
fixed points found in Sec.~\ref{subsec:Second-order-perturbation-theory}.
Since the SU(N)-symmetric angular fixed point $-\frac{\pi}{4}$ is
unstable, the angular distribution remains there only if $\theta_{0}=-\frac{\pi}{4}$.
In all other cases, the angular distribution flows to either $\theta=-\pi/2$
or $0$, depending on the initial value of $\theta_{0}$. For $-3\pi/4<\theta_{0}<-\pi/4$,
the flow is towards $\theta=-\pi/2$, whereas angles in the complementary
region flow towards $\theta=0$. Singlets are formed throughout the
chain, with spins paired two-by-two, but with otherwise randomly distributed
positions and sizes. Extending the conventions of the spin-1 chain
\cite{Quito_PhysRevLett.115.167201}, we will name this a ``mesonic''
phase.

\begin{figure}[H]
\begin{centering}
\includegraphics[scale=0.43]{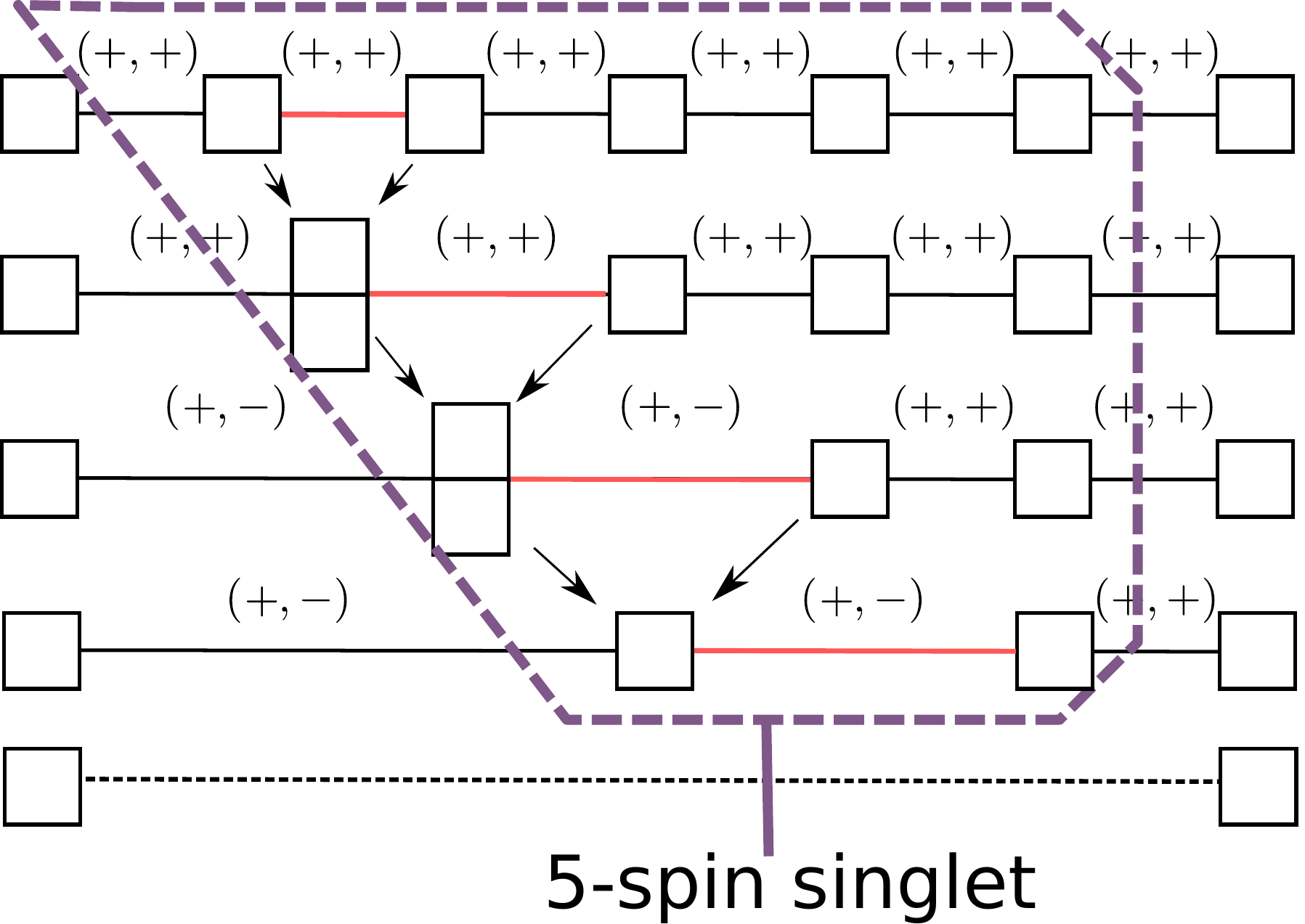}
\par\end{centering}
\caption{From top to bottom, a possible decimation route leading to a 5-site
singlet in an SO(5) chain. In red, we highlight the bond that is being
decimated at a given step. Above each bond is the sign of its respective
couplings $K_{i}^{\left(1\right)}$ and $K_{i}^{\left(2\right)}$.
More than the numeric renormalization of the constants, the sign flips
are crucial to follow the representation flow. In this example the
net result of the first three steps is to flip a sign of one of the
couplings, such that the pair of fundamental representations form
a singlet ground state in the last step. \label{fig:SUN_baryons}}
\end{figure}

In contrast to the case above, if the initial angle lies in the region
$\theta_{AKLT}<\theta_{0}<\frac{3\pi}{4}$, first-order decimations
will happen and the distribution of group representations will also
flow. This is analogous to what happens in the SDRG flow when the
couplings are random in sign in Heisenberg SU(2) spin chains, with
the important difference that here only a few representations will
enter the flow. As a consequence of the limited number of representations,
we are \emph{guaranteed} to obtain a singlet after a finite number
of steps. The remaining question is what is the character of the ground
states as well as of its low-lying excitations? The answer is that
the ground state is formed by a collection of singlets formed by $\mathrm{N}=5$
or any integer multiple of $5$ spins. One possible decimation route
is exemplified in Fig.~\ref{fig:SUN_baryons}, for a five-spin singlet.
In that Figure, we show explicitly the signs of the couplings $\left(K_{i}^{\left(1\right)},K_{i}^{\left(2\right)}\right)$
of each bond. In each decimation, Fig.~\ref{fig:All-decimations-SO(5)}
has been used to determine whether the signs of neighboring bonds
change or remain the same. Other cases can be worked at will, all
of them yielding (integer$\times$5)-site singlets. This phase will
be called here a ``baryonic'' phase.\cite{Quito_PhysRevLett.115.167201}

To characterize the mesonic and baryonic phases thermodynamically,
we have to determine the exponent $\psi$. For that, one first notices
that only two types of decimation processes occur, those of first
or second order. The analysis was carried out in Refs.~\onlinecite{Damle2002,PhysRevB.70.180401}
and will not be repeated here. Briefly, in the limit of wide distributions,
the numerical prefactors of the renormalized couplings can be safely
neglected. As a result, because of the multiplicative structure of
Eq.~(\ref{eq:K-order2}), only second-order decimations are effective
at lowering the energy scale. It follows that, if asymptotically $p$
is the fraction of second-order decimations, then 
\begin{equation}
\psi=\frac{1}{1+\frac{1}{p}}.
\end{equation}
In the mesonic phase, there are only second-order processes so $p=1$
and $\psi_{M}=1/2$. On the other hand, in the baryonic phase, $p=1/4$
and $\psi_{B}=1/5$.

\begin{figure}
\includegraphics[width=1\columnwidth]{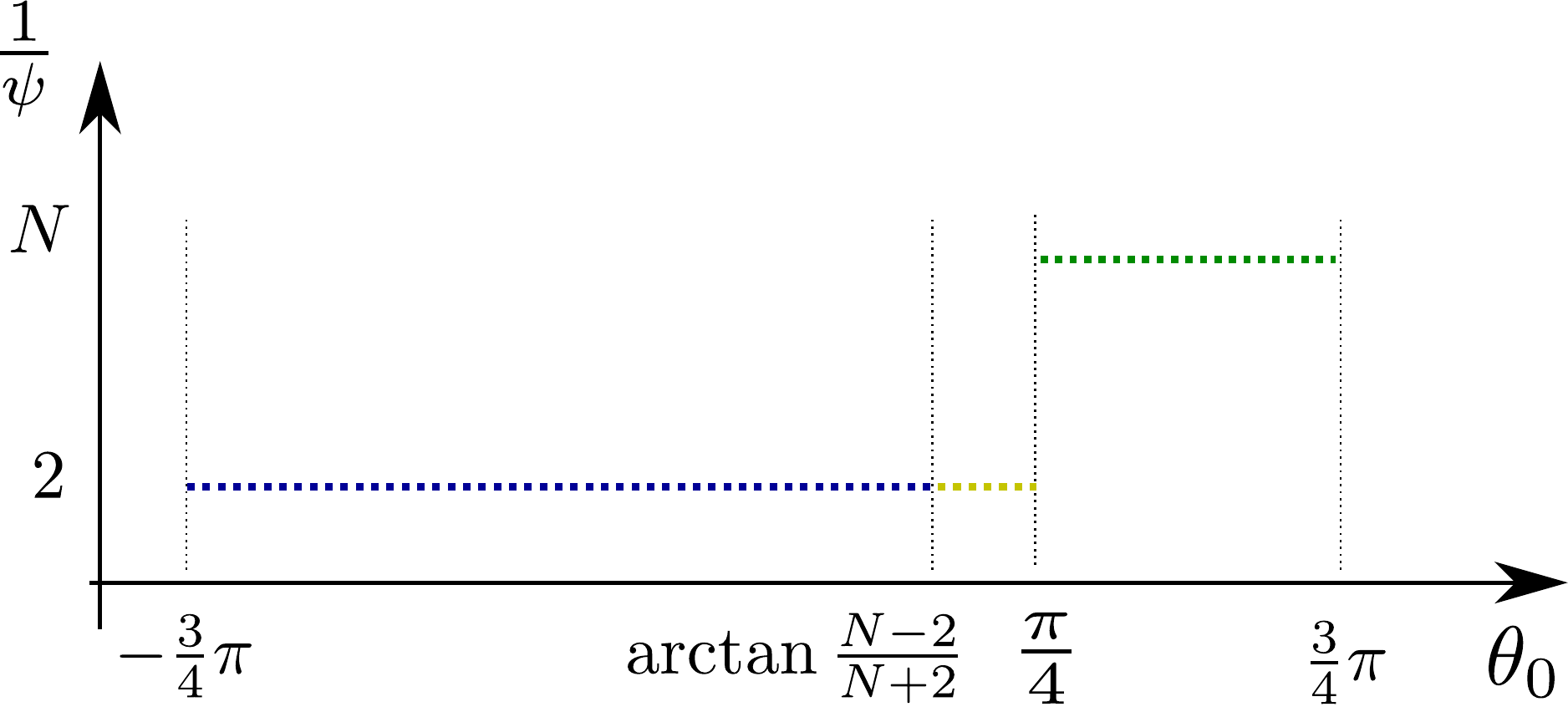}

\caption{Exponent $\psi$ as a function of the initial angle $\theta_{0}$,
defined in Eq.~(\ref{eq:angle-variables}), for SO(N)-symmetric Hamiltonians.
The blue region is the mesonic phase whereas the green one is the
baryonic phase. The yellow region $\theta_{AKLT}<\theta_{0}<\pi/4$
cannot be analyzed with the current method for even N, but should
be regarded as blue (mesonic) for odd N. \label{fig:Exponent-psi}}
\end{figure}

The SO(5) case can now be extrapolated to SO(N) for any odd value
of N, as we have carefully checked numerically for the lowest odd
N values. The RG structure is very similar, with the only mentioned
difference that the number of representations involved, in addition
to the singlet, is larger. Thus, the ground state for SO(N), odd N,
will be a collection of singlets made of either pairs of spins (mesonic
phase), or multiples of N spins (baryonic phase). The mesonic phase
is characterized by $\psi_{M}=1/2$, as only second-order processes
occur. In contrast, in the baryonic phase, the asymptotic probability
of the latter processes is $p=1/\left(\mathrm{N}-1\right)$ and thus
$\psi_{B}=1/\mathrm{N}$.~\cite{Damle2002,PhysRevB.70.180401} Figure~\ref{fig:Exponent-psi}
shows the value of the $\psi$ exponent as a function of the initial
angle $\theta_{0}$. The mesonic phase is identified by the blue color
whereas the baryonic one is green. The yellow region for $\arctan\left(\frac{\mathrm{N}-2}{\mathrm{N}+2}\right)<\theta_{0}<\pi/4$
also belongs to the mesonic phase. As we will see next, the diagram
is also valid for even values of N, although in that case the yellow
region cannot be analyzed with the current approach.

\subsection{SO(N) for even values of N}

For even N, we start with SO(4). Even though it is the yet unexplored
lowest-N case where our tools can be applied, it is very special because
its Clebsch-Gordan series for the product of two fundamental representations
has an additional term not present for any other value of N. As mentioned
in connection with Eq.~(\ref{eq:3terms}), the product of two fundamental
representations of SO(N) generically yields three terms.\cite{Iachello_book}
For SO(4), however, a fourth term is present. The proof of this statement
can be found in Ref.~\onlinecite{Iachello_book}. The additional
term in the Clebsch-Gordan series for SO(4) affects the form of the
most generic Hamiltonian for a pair of spins. Specifically, for pair
of spins, one can have

\begin{equation}
\mathcal{H}_{SO(4)}=J\mathbf{L}_{1}\cdot\mathbf{L}_{2}+D\left(\mathbf{L}_{1}\cdot\mathbf{L}_{2}\right)^{2}+F\epsilon^{ijkl}L_{1}^{ij}L_{2}^{kl}\label{eq:Hamilt_SO(4)}
\end{equation}
where $\epsilon^{ijkl}$ is the totally antisymmetric tensor ($i,j,k,l=1,\ldots,4$)
and the term proportional to $F$ is not allowed for $N\ne4$. The
spectrum of Eq.~(\ref{eq:Hamilt_SO(4)}) and the degeneracy of each
level are listed on the top of Table~\ref{tab:Spectrum-SO(4)}.

Since the group SO(4) is isomorphic to SU(2)$\otimes$SU(2), an equivalent
way of thinking about the SO(4) Hamiltonian is in terms of two spins-1/2
per site. Let us make the connection between the two languages. Denoting
the two spin-1/2 degrees of freedom by $\mathbf{S}$ and $\mathbf{T}$,
the most general two-site SU(2)$\otimes$SU(2)-invariant Hamiltonian
has the following form

\begin{equation}
\mathcal{H}_{KK}=B_{0}+B_{1}\mathbf{S}_{1}\cdot\mathbf{S}_{2}+B_{2}\mathbf{T}_{1}\cdot\mathbf{T}_{2}+B_{12}\left(\mathbf{S}_{1}\cdot\mathbf{S}_{2}\right)\left(\mathbf{T}_{1}\cdot\mathbf{T}_{2}\right).\label{eq:SO(4)-Hamit-2}
\end{equation}
This is the well-known Kugel-Khomskii model.\cite{Kugel1982} Two
of its good quantum numbers are associated with $\mathbf{T}_{T}^{2}=\left(\mathbf{T}_{1}+\mathbf{T}_{2}\right)^{2}$
and $\mathbf{S}_{T}^{2}=\left(\mathbf{S}_{1}+\mathbf{S}_{2}\right)^{2}$
and can be used to label the eigenstates of the Hamiltonian. The energy
levels and the corresponding quantum numbers associated with $\mathbf{T}_{T}^{2}=T_{T}\left(T_{T}+1\right)$
and $\mathbf{S}_{T}^{2}=S_{T}\left(S_{T}+1\right)$ are represented
on the right-hand side of Table~\ref{tab:Spectrum-SO(4)}. The equivalence
between Eqs.~(\ref{eq:Hamilt_SO(4)}) and (\ref{eq:SO(4)-Hamit-2})
is obtained with the following relations

\begin{align}
B_{0} & =\frac{3}{2}D,\\
B_{1} & =2\left(J-D-4F\right),\\
B_{2} & =2\left(J-D+4F\right),\\
B_{12} & =8D.
\end{align}

In order to make the SO(4)-symmetric Hamiltonian similar to the other
SO(2N) models, we will set $F=0$ in Eq.~(\ref{eq:Hamilt_SO(4)}).
The case where $F$ is non-zero leads to an SDRG flow which cannot
be treated with the approach described in this paper. Setting $F=0$
is equivalent to setting $B_{1}=B_{2}$ in (\ref{eq:SO(4)-Hamit-2}).
In this case, there is an additional $\mathbb{Z}_{2}$ symmetry related
to the exchange $\mathbf{S}\rightleftarrows\mathbf{T}$. With this
choice, the two triplet representations of SO(4) become degenerate
(see Table \ref{tab:Spectrum-SO(4)}), and the RG structure becomes
identical to any other SO(2N) model. 

\begin{table}
\begin{centering}
\begin{tabular}{|c|c|}
\hline 
degeneracy & $E$\tabularnewline
\hline 
\hline 
1 & $-3J+9D$\tabularnewline
\hline 
3 & $-J+D-8F$\tabularnewline
\hline 
3 & $-J+D+8F$\tabularnewline
\hline 
9 & $J+D$\tabularnewline
\hline 
\end{tabular}
\par\end{centering}
\begin{centering}
\begin{tabular}{|c|c|c|c|}
\hline 
degeneracy & $S_{T}$ & $T_{T}$ & $E$\tabularnewline
\hline 
\hline 
1 & 0 & 0 & $-\frac{3}{4}\left(B_{1}+B_{2}\right)+\frac{9}{16}B_{12}+B_{0}$\tabularnewline
\hline 
3 & 1 & 0 & $\frac{1}{4}\left(B_{1}-3B_{2}\right)-\frac{3}{16}B_{12}+B_{0}$\tabularnewline
\hline 
3 & 0 & 1 & $\frac{1}{4}\left(-3B_{1}+B_{2}\right)-\frac{3}{16}B_{12}+B_{0}$\tabularnewline
\hline 
9 & 1 & 1 & $\frac{1}{4}\left(B_{1}+B_{2}\right)+\frac{1}{16}B_{12}+B_{0}$\tabularnewline
\hline 
\end{tabular}
\par\end{centering}
\caption{Spectrum of the most general SO(4) Hamiltonian following the convention
of Eq.~(\ref{eq:Hamilt_SO(4)}) (top) and Eq.~(\ref{eq:SO(4)-Hamit-2})
(bottom).\label{tab:Spectrum-SO(4)}}
\end{table}

\begin{center}
\begin{figure}
\begin{centering}
\includegraphics[width=0.85\columnwidth]{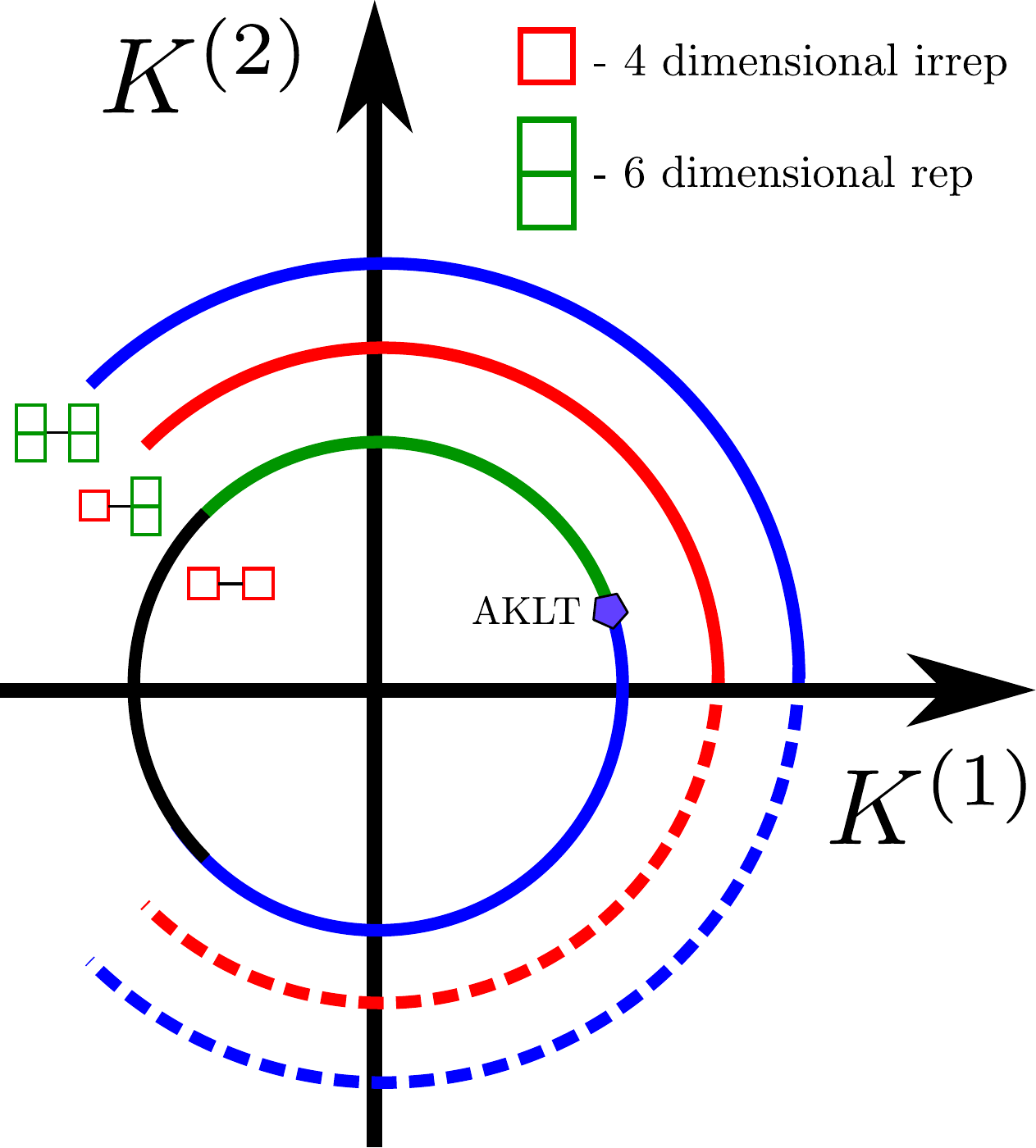}
\par\end{centering}
\caption{(Color online) Ground state structure of the SO(4) two-site problem
as a function of the angle $\theta$, with different colors representing
the distinct multiplets. Dashed lines correspond to different representations
of the same dimension as their corresponding color. The singlet state
is represented in blue while the fundamental representation $\left[1,0\right]$
(single box, in Young-tableau notation) in shown in red and the 6-dimensional
representation $\left[1,1\right]$ (two boxes) in green (cf. Fig.~\ref{fig:levels-so5}).
Unlike in the odd-$N$ case, the two-site gap closes at $\theta=0$,
which makes the RG flow ill-defined, except when the fundamental representation
of the group is realized on both sites (inner circle).\textcolor{blue}{{}
}\label{fig:levels-so4}}
\end{figure}
\par\end{center}

In general, the RG flow for SO(4) is almost identical to the case
we described for odd N, with one remarkable difference: the low-energy
physics of the region between the AKLT point $\theta_{0}=\arctan\frac{1}{3}$
and $\theta_{0}=\frac{\pi}{4}$ is ill-controlled within the SDRG
framework we are describing here. In that region, the initial RG structure
is very similar to the case described for odd $N$. The representation
$\left[1,1\right]$ is generated and SDRG rules that include such
representation are also necessary. As the SDRG proceeds, the angle
distribution starts flowing to a delta function at $\theta=0$. The
two outermost circles of Fig.~\ref{fig:levels-so4} show that the
local two-site gap \emph{closes} at $\theta=0$ when a pair of sites
with representations $\left[1,1\right]$-$\left[1,0\right]$ or $\left[1,1\right]$-$\left[1,1\right]$
are coupled. Since a large local gap is required for the validity
of perturbation theory, this makes the SDRG flow ill-defined asymptotically.\textcolor{blue}{{}
}In order to probe the physics of this region one has to go beyond
the current SDRG framework, keeping more than one multiplet when a
pair of sites is decimated, in a similar fashion to what was done
in Refs.~\onlinecite{monthusetal97,monthusetal98} in a different
context. Again, this falls outside the scope of this work. 

Outside this problematic region, the SDRG flow has the same structure
as the odd-N case of the previous Section. There is a mesonic phase
for $-3\pi/4<\theta_{0}<\arctan\frac{1}{3}$ with two possible angular
fixed points and $\psi_{M}=1/2$. If $-3\pi/4<\theta_{0}<-\pi/4$,
the flow is towards $\theta=-\pi/2$, whereas the angular fixed point
is $\theta=0$ if $-\pi/4<\theta_{0}<\arctan\frac{1}{3}$. For $\pi/4<\theta_{0}<3\pi/4$,
the phase is baryonic (with singlets made out of $4k$ original spins),
the angular fixed point is $\theta=\pi/2$, and $\psi_{B}=1/4$.

\begin{figure}
\includegraphics[width=0.8\columnwidth]{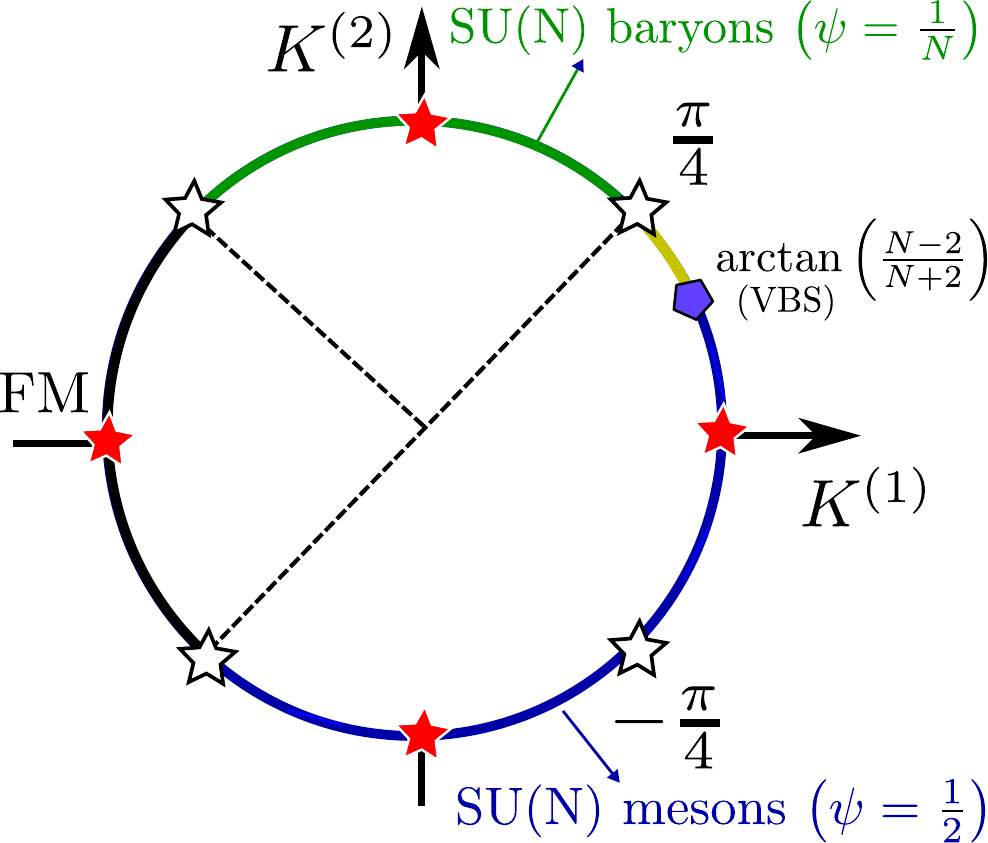}

\caption{The SO(N) spin chain phase diagram, reproduced from Ref.~\onlinecite{QuitoLopes2017PRL}.
Red stars correspond to stable SDRG fixed points, while white ones
correspond to the unstable fixed points. The point $\pi/4$ is SU(N)
symmetric, with the spins in the fundamental representation of the
group. The point $-\pi/4$ is also SU(N) symmetric, with the fundamental
and anti-fundamental representations on alternating sites. The black
arc denotes the ferromagnetic region, beyond the scope of this work.
The points $\pm3\pi/4$ fix the transition between the baryonic and
mesonic phases to the ferromagnetic region. The region between the
AKLT point $\left(\tan\theta=\frac{N-2}{N+2}\right)$ and the SU(N)-symmetric
point $\pi/4$ results either in an uncontrolled SDRG flow, for even
$N$, or for a basin of attraction equivalent to the blue region,
for odd $N$.\textcolor{blue}{{} }\label{fig:SO(N)-phase-diagram}}
\end{figure}

The SDRG flow and phase diagram for larger, even values of N is identical
to the one described for SO(4) although more representations are generated,
as shown in Fig.~\ref{fig:SO(N)-phase-diagram}. The region $\arctan\left(\frac{N-2}{N+2}\right)<\theta_{0}<\frac{\pi}{4}$
suffers the same problems as in SO(4) and cannot be properly treated
with the current SDRG scheme (yellow region of Fig.~\ref{fig:SO(N)-phase-diagram}).
The mesonic and baryonic phases are shown as blue and green in Fig.~\ref{fig:SO(N)-phase-diagram},
respectively. The former is characterized by $\psi_{M}=1/2$ whereas
the latter has $\psi_{B}=1/\mathrm{N}$. If we conventionalize that
for odd N, the yellow region has the same physics as the blue one,
Fig.~\ref{fig:SO(N)-phase-diagram} encapsulates the phase diagram
of all SO(N)-symmetric disordered spin chains.

\subsection{The Sp(N) Group}

\begin{figure}
\includegraphics[width=1\columnwidth]{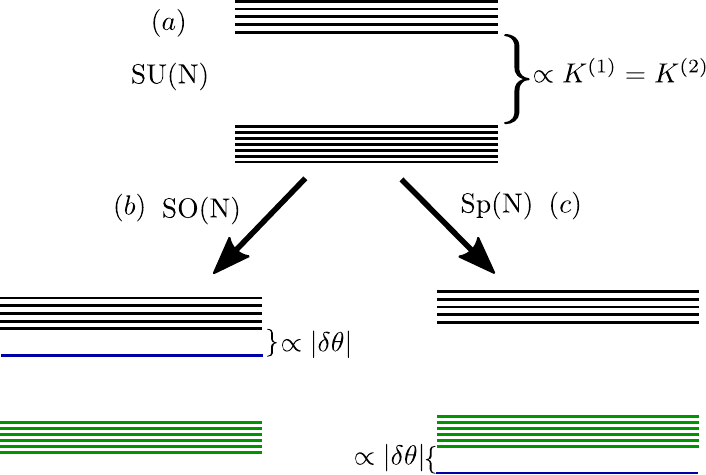}

\caption{Schematic energy levels close to the point $\frac{\pi}{4}$, in terms
of the angular deviation $\delta\theta=\left|\theta-\frac{\pi}{4}\right|\ll1$.
(a) At the point $\frac{\pi}{4}$, the Hamiltonian has explicit SU(N)
symmetry, and the degeneracies are $\frac{\mathrm{N}\left(\mathrm{N}-1\right)}{2}$
(ground multiplet) and $\frac{\mathrm{N}\left(\mathrm{N}+1\right)}{2}$
(excited multiplet). (b) Adding a small perturbation that breaks the
SU(N) symmetry into SO(N), the excited multiplet splits into levels
with a $\frac{\mathrm{N}\left(\mathrm{N}+1\right)}{2}-1$ degeneracy
and a singlet, while the low-energy multiplet remains the same (c)
Now, slightly breaking SU(N) into Sp(N). The lowest energy state is
a singlet (blue), separated from the first excited multiplet by a
small gap proportional to $\left|\delta\theta\right|$. \label{fig:SUN-break}}
\end{figure}

We now address Sp(N)-symmetric models. As Sp(N) is a subgroup of SU(N),
by fine-tuning the angle parameter the symmetry can be explicitly
enhanced, just as in the SO(N) case {[}see Eq.~(\ref{eq:AnisSpN}){]}.
The SU(N)-symmetric points are again at $\pm\frac{\pi}{4}$, and $\pm\frac{3\pi}{4}$.
There is, however, a remarkable difference between these high-symmetry
points, when compared to the SO(N) case. In SO(N) chains, the level
structure at the angle $\frac{\pi}{4}$ is such that two \emph{excited
states} of the two-site problem have the same energy. In Sp(N) systems,
on the other hand, the \emph{two lowest-lying multiplets }become degenerate
at the angle $\frac{\pi}{4}$. This can be predicted by directly looking
at the degeneracies of the Sp(N) multiplets, and comparing it to the
SU(N) degeneracies. The breaking of SU(N) into SO(N) or Sp(N) in terms
of energy levels of a two-site problem is shown in Fig.~\ref{fig:SUN-break}.
In contrast with SO(N), where the decimations are well-defined around
this high-symmetry point, in Sp(N) the local two-spin gap is proportional
to $\delta\theta=\left|\theta-\frac{\pi}{4}\right|$. If $\delta\theta$
is small, the SDRG as proposed in this work cannot be applied. 

\begin{figure}
\noindent \begin{centering}
\includegraphics[width=0.85\columnwidth]{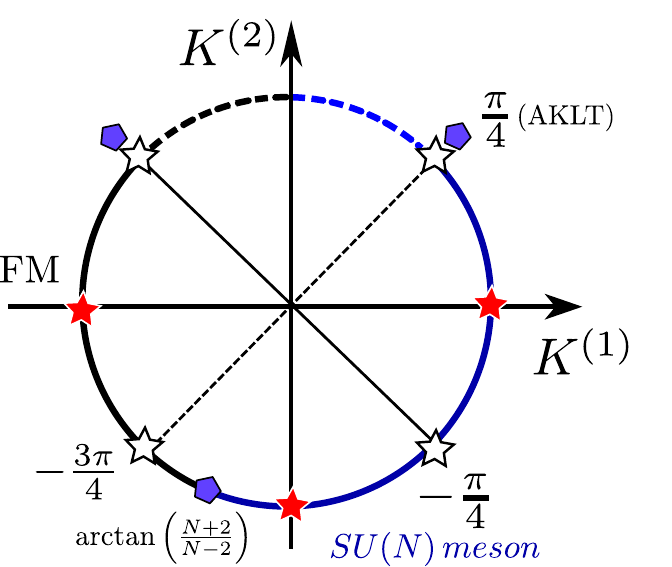}
\par\end{centering}
\caption{The phase diagram of Sp(N)-symmetric chains, with $N=4,6$. For $\arctan\frac{N+2}{N-2}<\theta<\frac{\pi}{2},$
the system is in a mesonic SU($N$) random singlet phase. Otherwise,
it is in a ferromagnetic like phase. The pentagons represent the angles
where two multiplets cross as ground states. The white stars correspond
to unstable angular fixed points, while the red ones are the stable
angular fixed points. Notice that there is no baryonic phase in Sp(N).\textcolor{blue}{{}
}Colors match their corresponding basins of attraction; the flow of
the dashed lines is described in the main text.\label{fig:Sp(N)-phase-diagram}}
\end{figure}

Still, there are regions of the phase diagram that can be safely analyzed
with our method. In order to characterize the AF phases, we again
assume that the initial angle is fixed. If the initial angle is $\arctan\left(\frac{N+2}{N-2}\right)<\theta_{0}<\frac{\pi}{4}$
(blue region of Fig.~\ref{fig:Sp(N)-phase-diagram}), and as long
as $\theta_{0}$ is far enough from the $\frac{\pi}{4}$ point such
that the SDRG is consistent, the distribution of angles will broaden
at early RG stages. After some transient, however, the distribution
converges to a delta function at either $0$ or $-\frac{\pi}{2}$,
which are the only stable angular fixed points in this region (see
Fig.~\ref{fig:SUN-break}), characterized by $\psi_{M}=1/2$. Again,
just like in the SO(N) case, the point $-\frac{\pi}{4}$ has a higher
SU(N) symmetry and corresponds to an unstable angular fixed point.
The basins of attraction of $\theta=0$ and $\theta=-\frac{\pi}{2}$
are $-\frac{\pi}{4}<\theta_{0}<\frac{\pi}{4}$ and $\arctan\left(\frac{N+2}{N-2}\right)<\theta_{0}<-\frac{\pi}{4}$,
respectively.

The most striking difference between SO(N) and Sp(N) chains appears
when the initial angle is in the range $\frac{\pi}{4}<\theta_{0}<\frac{3\pi}{4}$
(dashed line of Fig.~\ref{fig:Sp(N)-phase-diagram}). First, since
there is no SU(N)-symmetric point inside this region, the shortcut
we used to derive the pre-factors of the SDRG equations cannot be
used. The generic rules for first-order decimations {[}Eq.~(\ref{eq:K-order-1}){]}
are still valid, but, in order to determine the pre-factors, the Sp(N)
tensors have to be constructed explicitly on a case-by-case basis.
Determining these factors is mandatory to follow the SDRG flow, particularly
since under certain conditions, our first-order RG rules become ill-defined
when the proportionality constants vanish (see Section\textcolor{blue}{~\ref{subsec:First-order-perturbation-theory-gen}}).
To see this concretely, let us consider the cases of Sp(4) and Sp(6)
as examples, and discuss the general features that are expected to
appear for larger N. For Sp(4) the coupling $K^{\left(2\right)}$
is renormalized to zero due to the structure of the Clebsch-Gordan
series (case (i) of Section\textcolor{blue}{~\ref{subsec:First-order-perturbation-theory-gen}}).
For $\frac{\pi}{4}<\theta_{0}<\frac{3\pi}{4}$, the following Clebsch-Gordan
series are relevant~\cite{FegerTables}

\begin{align}
4\otimes4 & =1\oplus5\oplus10\nonumber \\
4\otimes5 & =4\oplus16\nonumber \\
5\otimes5 & =1\oplus10\oplus14\nonumber \\
5\otimes10 & =5\oplus10\oplus35,
\end{align}
where we labeled the representations by their dimensions. Recall that
for a pair of representations $\Upsilon$ and $\Upsilon'$, $P_{\Upsilon}T^{\Upsilon'}P_{\Upsilon}\ne0$,
$P_{\Upsilon}$ being a projection operator onto representation $\Upsilon$,
only if $\Upsilon'$ belongs to the Clebsch-Gordan series of $\Upsilon\otimes\Upsilon$.
Starting with the fundamental representation of dimension 4 on each
site, initial RG steps generate the 5-dimensional representations
via first-order decimations. After some steps, unavoidably, a decimation
of a 5-dimensional representation coupled to a 4-dimensional representation
happens. Let us label the Sp(4) tensors by $T^{10}$ (coupled by $K^{\left(1\right)}$)
and $T^{5}$ (coupled by $K^{\left(2\right)}$), using the short-hand
notation of labeling the representations by their dimension. Now,
since
\begin{align}
P_{5}T^{10}P_{5} & \ne0,\nonumber \\
P_{5}T^{5}P_{5} & =0,
\end{align}
the renormalized $K^{\left(2\right)}$, which is proportional to $P_{5}T^{5}P_{5}$,
is zero. A similar renormalization to zero has been found in SU(2)-symmetric
spin-$S$ chains with $S>1$.~\cite{quitoprb2016}. As a consequence,
the low-energy physics is dominated by $K^{\left(1\right)}$ only.
The initial sign of $K^{\left(1\right)}$ thus becomes crucial, and
a distinction has to be made depending on whether $\theta_{0}$ is
greater or smaller than $\frac{\pi}{2}$. If $\frac{\pi}{2}<\theta_{0}<\frac{3\pi}{4}$,
the renormalization projects $\left(K^{\left(1\right)},K^{\left(2\right)}\right)$
into the $K^{\left(1\right)}<0$ semi-axes, and the phase is ferromagnetic,
and thus outside the scope of our analysis. If, on the other hand,
the initial angle lies on the interval $\frac{\pi}{4}<\theta_{0}<\frac{\pi}{2}$,
the projection makes the flow identical to the one starting with $\theta=0$,
and the low-energy physics is again a RSP with $\psi_{M}=\frac{1}{2}$.

Similar reasoning can be applied to Sp(6). Here the important Clebsch-Gordan
series read~\cite{FegerTables}

\begin{align}
6\otimes6 & =1\oplus14\oplus21\nonumber \\
14\otimes6 & =6\oplus14'\oplus64\nonumber \\
14'\otimes6 & =14\oplus70\nonumber \\
14\otimes14' & =6\oplus64\oplus126\nonumber \\
14\otimes14 & =1\oplus14\oplus21\oplus70\oplus90\nonumber \\
14'\otimes14' & =1\oplus21\oplus84\oplus90\nonumber \\
14'\otimes21 & =14'\oplus64\oplus216.
\end{align}
Notice that the representations $14'$ and $14$ are different, even
though they have the same dimension. The tensors of interest are $T^{14}$
(coupled by $K^{\left(1\right)}$) and $T^{21}$ (coupled by $K^{\left(2\right)}$).
In the region of interest, initial RG steps generate representations
of dimension $14$, which, when coupled to the fundamental representation,
enforces $14'$ as the ground state. From the Clebsch-Gordan series
above, $P_{14'}T^{14}P_{14'}=0$, while $P_{14'}T^{21}P_{14'}\ne0$. 

We have checked that for Sp(8) and Sp(10) these renormalizations to
zero \emph{do not} appear, which means that these features are most
likely a property present for low N only. A different issue arises,
however. By a similar analysis of the Clesbch-Gordan series,~\cite{FegerTables}
we find that a large number of representations are generated in early
RG steps, as opposed to the SO(N) flow, where the number of representations
appearing when $\frac{\pi}{4}<\theta_{0}<\frac{3\pi}{4}$ is always
$\text{int}\left(\mathrm{N}/2\right)$. At this stage, the Sp(N) problem
might lead then to either a ferromagnetic phase, or a so-called large
spin phase (LSP),\cite{westerbergetal} also common in disordered
spin chains. Another possibility is that these large representations
disappear at low energies. The only way to see which one actually
happens is to construct the tensors from their definition (Eq.~(\ref{eq:tensor_def}))
for all these representations, as well as the pre-factor of the RG
rules (Eqs.~(\ref{eq:K-order-1}) and~(\ref{eq:K-order2})), a very
challenging task. Physically, since the most relevant Sp(N) cases
are the ones with small $N$, we will not pursue a further analysis
here.

\section{Emergent SU(N) Symmetry \label{sec:Emergent-Symmetries}}

In the previous Sections, we characterized the SDRG flow by determining
the AF phases as well as their exponent $\psi$. A more subtle feature
of the RSPs displayed above is the emergence of SU(N) symmetries.
In this Section, we show the mechanism responsible for the symmetry
enhancement. An overall explanation has been given in Ref.~\onlinecite{QuitoLopes2017PRL},
and here we complement it with further details.

We would like to emphasize the generality of this result. Notice that
we have studied all the random antiferromagnetic spin chain models
invariant under transformations of the semi-simple Lie groups, Sp($N$),
SO($N$), and, consequently, the SU($N$). We have focused, however,
only on the case in which the spins are represented only by totally
antisymmetric representations of the group.

Let us start with the SO(N) case. The scalar operators that constitute
Hamiltonian (\ref{eq:HamiltSON}) are formed out of tensor operators
$T_{\upsilon}^{\left[1,1\right]}$ and $T_{\upsilon}^{\left[2,0\right]}$.
These operators have a physical interpretation similar to the vector
and quadrupolar operators in SU(2),~\cite{Quito_PhysRevLett.115.167201}
and their response functions are not expected to be generically the
same. Indeed, thinking of the SO(N) problem as an anisotropic SU(N)
model, the uniform susceptibilities associated with the SU(N) generators
are
\begin{equation}
\chi_{\mu}=\beta\left(\left\langle \Lambda_{T}^{\mu}\right\rangle ^{2}-\left\langle \left(\Lambda_{T}^{\mu}\right)^{2}\right\rangle \right).
\end{equation}
If the Hamiltonian of the problem displays an SO(N)-preserving SU(N)
anisotropy, no reason a priori exists to expect the responses involving
the $\Lambda^{\mu}$ chosen from any of the $N\left(N-1\right)/2$
SU(N) purely imaginary generators {[}$\sim T_{\upsilon}^{\left[1,1\right]}$
in SO(N){]} to be the equal to those of the remaining $N\left(N+1\right)/2-1$
SU(N) purely real ones {[}$\sim T_{\upsilon}^{\left[2,0\right]}$
in SO(N){]}. Yet, as we explain below, in the RSPs the singular behavior
is isotropic and equal to that of an SU($N$)-invariant system.

Our argument is supported by two key observations: (i) each SO(N)
representation that appears throughout the SDRG flow has an SU(N)
counterpart and (ii) the 2-site ground states at the stable angular
fixed points $0,-\pi/2$ ($+\pi/2$) are identical to those of the
SU(N) symmetric points $-\pi/4$ $\left(+\pi/4\right)$. The direct
consequence of points (i) and (ii) is that if the angular distribution
starts at the stable fixed points, the SDRG flow at low energies is
indistinguishable from the flow started at the SU(N) symmetric points,
for the following reasons. Point (i) guarantees that the representations
generated in the flows within the mesonic and baryonic phases always
find counterparts in the SU(N) representation spectrum, so any remaining
non-decimated spin at low-energies still defines an object transforming
in the full SU(N) group. Furthermore, through point (ii), the local
two-spin gaps never close for angles between  $0$ and $-\pi/2$,
which includes the SU(N)-symmetric mesonic point $-\pi/4$, or between
the SU(N)-invariant baryonic point $\pi/4$ and $\pi/2$. Adding to
this the fact that the unstable SU(N)-symmetric points are contained
in the corresponding mesonic/baryonic AF basins of attraction, one
finally realizes that all the two-site ground multiplets generated
throughout the flow are SU(N) invariant. In sum, points (i) and (ii)
guarantee that both the ground state and the collection of free spins
at finite low-energies are composed of SU(N)-invariant objects. A
difference between the flows at the SU(N)-symmetric points and those
of the stable angular fixed points does remain. It lies in the fact
that the decimation rules for the radii $r_{i}$ have distinct pre-factors
depending on the bond angle. Since the radial disorder grows without
bounds in RSPs, however, the pre-factors are asymptotically irrelevant.

Combining the points above, we conclude that the ground state of the
system is an SU(N)-invariant state composed of a collection of SU(N)
singlets, while the low-energy physics of the chain is governed by
free spins in SU(N) anti-symmetric representations. From this, thermodynamic
quantities, such as the magnetic susceptibilities, follow immediately.
The calculation of the magnetic susceptibility $\chi_{\mu}$ for a
single free spin $\Lambda^{\mu}$ gives Curie's law $\chi_{\mu}^{free}\sim T^{-1}$,
independently of $\mu$. The total susceptibility is then obtained
by multiplying by the density of free spins at energy scale $\Omega=T$,
$\chi_{\mu}\left(T\right)^{-1}\sim TL_{T}\sim T\left|\ln T\right|^{1/\psi}$,
which is controlled by the universal exponent $\psi$.~\cite{QuitoLopes2017PRL}.
The impact of the distinct pre-factors of the SDRG decimation equations
is only in the non-universal behavior of the pre-factors, but not
in the universal exponents.

Both RSPs are then comprised of collections of completely frozen pairs
or $k$N-tuples of spins and low-energy free spin excitations which
actually transform as irreducible representations of SU(N). As discussed
in Sec.~\ref{sec:subgroups-1-1-1} we make an analogy with quantum
chromodynamics and we call the two AF phases mesonic or baryonic.
If the ground state is a collection of two-site singlets, we have
the\emph{ mesonic RSP,} with tunneling exponent is $\psi_{M}=1/2$.
For $\mathrm{N}=2$ , this is just the standard RSP phase for XXZ
spin-1/2 chains, that are SU(2) anisotropic, but which indeed display
emergent SU(2) symmetry at low energies.~\cite{fishertransising2,fisher94-xxz}
If, on the other hand, the phase is characterized by a collection
of singlets formed out of multiples of N spins, then we have the \emph{baryonic
RSP}. This phase has a tunneling exponent $\psi_{B}=1/\mathrm{N}$.
Crucially, these tunneling exponents are indeed the same ones found
previously by two of us in the context of SU(N)-symmetric disordered
spin chains.~\cite{PhysRevB.70.180401}

While the baryonic RSP is generically attainable for SO(N) Hamiltonians,
we see that it is not in the Sp(N) case, which displays only mesonic
phases (due to the adiabaticity argument for $\theta=-\pi/4$, see
Fig.~\ref{fig:Sp(N)-phase-diagram}). This a striking distinction
arising from the fact that the other SU(N)-symmetric point, $\theta=+\pi/4$,
is located now at a ground multiplet degeneracy point (AKLT) (again,
see Fig.~\ref{fig:Sp(N)-phase-diagram}). At that point, our SDRG
rules break down and a more refined analysis must be made on a case-by-case
basis.

Finally, another hallmark of the emergent symmetry phenomenon is the
ground-state spin-spin correlation function $C_{i,j}^{\mu}=\left\langle \Lambda_{i}^{\mu}\Lambda_{j}^{\mu}\right\rangle $
($\mu$ label not summed). As we have just shown, the ground state
of the anisotropic model is identical to the isotropic one in the
RSP within the approximation of the SDRG method. Since the singular
behavior is captured exactly (asymptotically) by the SDRG method,
it is then a straightforward conclusion that in the RSPs here reported
$C_{i,j}^{\mu}=C_{i,j}$ as far as the singular behavior is concerned.
Therefore, the mean $\overline{C_{i,j}^{\mu}}$ and typical $C_{i,j}^{\mu,\text{typ}}$
values of the correlation function are those of the SU($N$) symmetric
models, already reported in the literature.~\cite{PhysRevB.70.180401,xavier-etal-prb18}
Thus,
\begin{equation}
\overline{C_{i,j}^{\mu}}\sim\left|i-j\right|^{-\eta_{\mu}},\label{eq:mean-C}
\end{equation}
 with universal exponent $\eta_{\mu}=\eta=2$ (both in the mesonic
and baryonic phases), and 
\begin{equation}
C_{i,j}^{\mu,\text{typ}}\sim\exp-\left|\frac{i-j}{\xi}\right|^{\psi},\label{eq:typ-C}
\end{equation}
with the universal tunneling exponent $\psi$ (which is either $\psi_{M}=\frac{1}{2}$
or $\psi_{B}=\frac{1}{N}$). Here $\xi$ is a non-universal length
scale of order of the crossover length between the clean and the infinite-randomness
fixed points.

\section{RSP signatures in Higher-order susceptibilities \label{sec:non-linear-suscep}}

A naturally relevant question regards how to detect signatures of
symmetry emergence. In Section~\ref{sec:Emergent-Symmetries}, we
mentioned that such signatures can be seen in the linear susceptibilities
of $T_{\upsilon}^{\left[1,1\right]}$ and $T_{\upsilon}^{\left[2,0\right]}$
operators. These tensors present the same low-temperature dependence,
even though this is not obvious \emph{a priori,} given the anisotropy
of the underlying Hamiltonian. Since the possible realizations of
SO(N) chains have very different microscopic origins, it is difficult
to give a generic prescription of how to access such susceptibilities,
i.e. one that is valid for any value of N.

To make progress, we attempt to draw inspiration from a concrete case:
$N=3$ (the spin-1 chain). In this case, the $T_{\upsilon}^{\left[1,1\right]}$
operators are the usual spin-1 (vector) operators $S_{x}$, $S_{y}$
and $S_{z}$ and their corresponding linear susceptibilities are the
usual magnetic susceptibilities. The $T_{\upsilon}^{\left[2,0\right]}$
operators are the spin-1 quadrupolar operators $\left(3S_{z}^{2}-2\right)/\sqrt{3}$,
$S_{x}^{2}-S_{y}^{2}$, $S_{x}S_{y}+S_{y}S_{x}$, $S_{x}S_{z}+S_{z}S_{x}$,
and $S_{y}S_{z}+S_{z}S_{y}$ \cite{Quito_PhysRevLett.115.167201}.
Since these $T_{\upsilon}^{\left[2,0\right]}$ operators involve products
of the vector operators, it is suggestive that higher-order, \emph{non-linear},
susceptibilities of $T_{\upsilon}^{\left[1,1\right]}$ might serve
as a window to study \emph{linear} susceptibilities of $T_{\upsilon}^{\left[2,0\right]}$,
thus serving as a good route to distinguish RSPs with and without
symmetry enhancement. For example, for $N=3$, inspection of the first
quadrupolar operator ($\sim S_{z}^{2}+\mathrm{const.}$) naively suggests
that the first non-linear magnetic susceptibilities might come in
handy. 

So we take here a position: in the general scenario, we assume that
$T_{\upsilon}^{\left[1,1\right]}$ susceptibilities are easier to
access and ask whether we could use non-linear susceptibilities of
$T_{\upsilon}^{\left[1,1\right]}$ to probe $T_{\upsilon}^{\left[2,0\right]}$
and, therefore, the symmetry enhancement. We show, however, that the
structure of the RSPs is such that this is, in fact, incorrect. Non-linear
responses of $T_{\upsilon}^{\left[1,1\right]}$ operators show \emph{no
distinction} between phases with and without symmetry enhancement.

Let us denote by $\Lambda^{\mu}=\sum_{i=1}^{N_{s}}\Lambda_{i}^{\mu}$
the total value (summed over all $N_{s}$ sites) of the $\mu$-th
SU(N) generator ($\mu=1,\ldots,\mathrm{N}^{2}-1$). As explained before,
the first $\mathrm{N}\left(\mathrm{N}+1\right)/2$ of these are the
generators of SO(N) (the $T_{\upsilon}^{\left[1,1\right]}$ operators),
whereas the remainder ($\mu=\mathrm{N}\left(\mathrm{N}+1\right)/2+1,\ldots,\mathrm{N}^{2}-1$)
are the $T_{\upsilon}^{\left[2,0\right]}$ tensor operators of SO(N).
Coupling external fields to these quantities $H\to H+h_{\mu}\Lambda^{\mu}$,
the expressions for the linear and the first non-zero non-linear susceptibilities
can be obtained as

\begin{align}
\chi_{\mu}^{\left(1\right)} & =\left.\frac{\partial\left\langle \Lambda^{\mu}\right\rangle }{\partial h_{\mu}}\right|_{h_{\mu}\to0}=\frac{1}{T}\left[\left\langle \left(\Lambda^{\mu}\right)^{2}\right\rangle -\left\langle \Lambda^{\mu}\right\rangle ^{2}\right],\\
\chi_{\mu}^{\left(3\right)}= & \left.\frac{\partial^{3}\left\langle \Lambda^{\mu}\right\rangle }{\partial h_{\mu}^{3}}\right|_{h_{\mu}\to0}=\frac{1}{T^{3}}\left[\left\langle \left(\Lambda^{\mu}\right)^{4}\right\rangle -4\left\langle \Lambda^{\mu}\right\rangle \left\langle \left(\Lambda^{\mu}\right)^{3}\right\rangle \right.\nonumber \\
 & \left.-3\left\langle \left(\Lambda^{\mu}\right)^{2}\right\rangle ^{2}+12\left\langle \Lambda^{\mu}\right\rangle ^{2}\left\langle \left(\Lambda^{\mu}\right)^{2}\right\rangle -6\left\langle \Lambda^{\mu}\right\rangle ^{4}\right],
\end{align}
where $\left\langle \mathcal{O}\right\rangle $ represents the thermal
expectation value of $\mathcal{O}$. In the absence of symmetry breaking
$\left\langle \Lambda^{\mu}\right\rangle =0$ and the expressions
simplify to 

\begin{align}
T\chi_{\mu}^{\left(1\right)} & =\left\langle \left(\Lambda^{\mu}\right)^{2}\right\rangle ,\\
T^{3}\chi_{\mu}^{\left(3\right)} & =\left\langle \left(\Lambda^{\mu}\right)^{4}\right\rangle -3\left\langle \left(\Lambda^{\mu}\right)^{2}\right\rangle ^{2}\nonumber \\
 & =\left\langle \left(\Lambda^{\mu}\right)^{4}\right\rangle -3T^{2}\left[\chi_{\mu}^{\left(1\right)}\right]^{2}.
\end{align}

We want to write SDRG results for these quantities in the limit $h_{\alpha}\ll T$
in the various random singlet phases. Stopping the SDRG flow when
the largest coupling $\Omega$ reaches some low temperature $T$,
there are asymptotically two types of objects left: free spins, with
density $n\left(T\right)\sim1/L_{T}\sim\left|\ln T\right|^{-1/\psi}$,
and strongly bound SO(N) singlets, with density $\propto1-n\left(T\right)$.
The actual density of singlets depends on how many original spins
are required to form them. In the mesonic phases, this is $\left[1-n\left(T\right)\right]/2$.
In the baryonic ones, in which singlets are composed of $k\mathrm{N}$
original spins, it is $\left[1-n\left(T\right)\right]/\overline{k}\mathrm{N}$,
where $\overline{k}\gtrsim1$ is the average value of $k$. In general,
the linear susceptibilities can then be written as

\begin{equation}
T\chi_{\mu}^{\left(1\right)}\sim n\left(T\right)\left\langle \left(\Lambda_{i_{0}}^{\mu}\right)^{2}\right\rangle _{\text{free}}+\frac{1-n\left(T\right)}{C}\left\langle \left(\Lambda^{\mu}\right)^{2}\right\rangle _{\text{singlet}},\label{eq:linear_suscep}
\end{equation}
where $C=2$ or $\overline{k}\mathrm{N}$, whichever is the case,
and the expectation values should be calculated in the ground multiplets,
either a free spin or a random singlet. In this equation, $i_{0}$
labels an arbitrary free spin site and $\Lambda^{\mu}=\sum_{i=1}^{k\mathrm{N}}\Lambda_{i}^{\mu}$,
where the sum is over all the $k\mathrm{N}$ spins within a singlet.
The expectation values of the free spins are independent of $i_{0}$.
Analogously,
\begin{eqnarray}
T^{3}\chi_{\mu}^{\left(3\right)}+3T^{2}\left[\chi_{\mu}^{\left(1\right)}\right]^{2} & \sim & n\left(T\right)\left\langle \left(\Lambda_{i_{0}}^{\mu}\right)^{4}\right\rangle _{\text{free}}+\nonumber \\
 & + & \frac{1-n\left(T\right)}{C}\left\langle \left(\Lambda^{\mu}\right)^{4}\right\rangle _{\text{singlet}}.\label{eq:nonlinear_suscep}
\end{eqnarray}

Let us now analyze separately the cases of SO(N) $T_{\upsilon}^{\left[1,1\right]}$
and $T_{\upsilon}^{\left[2,0\right]}$ operators. For the sake of
clarity, we will use labels $\mu\to\alpha\in\left[1,\ldots,\mathrm{N}^{2}-1\right]$
for the former and $\mu\to\beta\in\left[\mathrm{N}\left(\mathrm{N}+1\right)/2+1,\ldots,\mathrm{N}^{2}-1\right]$
for the latter. Now, since the SO(N) singlets are annihilated by the
total SO(N) generators, $\left\langle \left(\Lambda^{\alpha}\right)^{2}\right\rangle _{\text{singlet}}=\left\langle \left(\Lambda^{\alpha}\right)^{4}\right\rangle _{\text{singlet}}=0$.
Thus, for the $T_{\upsilon}^{\left[1,1\right]}$ operators,
\begin{equation}
T\chi_{\alpha}^{\left(1\right)}\sim n\left(T\right)\left\langle \left(\Lambda_{i_{0}}^{\alpha}\right)^{2}\right\rangle _{\text{free}},\label{eq:linear_suscep2}
\end{equation}
and

\begin{eqnarray}
T^{3}\chi_{\alpha}^{\left(3\right)}+3T^{2}\left[\chi_{\alpha}^{\left(1\right)}\right]^{2} & \sim & n\left(T\right)\left\langle \left(\Lambda_{i_{0}}^{\alpha}\right)^{4}\right\rangle _{\text{free}}.\label{eq:non-linear-suscep2}
\end{eqnarray}
This equation is the main finding of this section. Eq.~(\ref{eq:non-linear-suscep2})
is completely expressed in terms of SO(N) labels $\alpha$ and is,
in general, unrelated to $\chi_{\beta}^{\left(1\right)}$. This expression
is the same for arbitrary RSPs \emph{independently of symmetry emergence}
(see especially the following Section~\ref{sec:When-the-emergent}).
In summary, non-linear susceptibilities are \emph{not} useful in assessing
whether or not an RSP displays symmetry emergence or not.

The result above is in strong contrast to the linear susceptibility
of tensor SO(N) operators $\chi_{\beta}^{\left(1\right)}$, which
is, indeed a probe of symmetry enhancement as shown before. Let us
briefly revisit the argument for completeness. If the SO(N) singlet
is also an SU(N) singlet, a \emph{necessary} ingredient for the emergent
symmetry, then the linear susceptibilities for tensors $T_{\upsilon}^{\left[2,0\right]}$
also obey

\begin{equation}
\left\langle \left(\Lambda^{\beta}\right)^{2}\right\rangle _{\text{singlet}}\underbrace{=}_{\text{ES}}0,\label{eq:singletcontr}
\end{equation}
where ``ES'' indicates the presence of an emergent symmetry. This
is the case since the $\Lambda^{\beta}$ are also SU(N) generators
and, therefore, annihilate SU(N) singlets. In summary, neglecting
non-universal prefactors,
\begin{align}
\chi_{\beta}^{\left(1\right)}\sim & \begin{cases}
\frac{n\left(T\right)}{T}\left\langle \left(\Lambda_{i_{0}}^{\beta}\right)^{2}\right\rangle _{\text{free}}\sim\frac{1}{T\left|\ln T\right|^{1/\psi}} & ,\text{ if ES}\\
\frac{1-n\left(T\right)}{CT}\left\langle \left(\Lambda_{i_{0}}^{\beta}\right)^{2}\right\rangle _{\text{singlet}}\sim\frac{1}{T} & ,\text{ otherwise,}
\end{cases},\label{eq:linear-suscep_tensor}
\end{align}
indeed distinct from what we see for $\chi_{\alpha}^{\left(1\right)}$
in Eq.~(\ref{eq:non-linear-suscep2}), which is always true and $\sim n\left(T\right)/T$.

\section{An RSP without emergent symmetry \label{sec:When-the-emergent}}

So far we have focused on a fairly general class of models with manifest
SO(N) symmetry in which RSPs displaying a larger SU(N) symmetry emerges
at low energies. The question then arises: is there a counterexample
to this situation, namely, a model with an RSP phase in which no larger
symmetry emerges? What physical consequences would follow in that
case? We will now show a specific model in which this does indeed
happen.

\begin{figure}
\includegraphics[width=1\columnwidth]{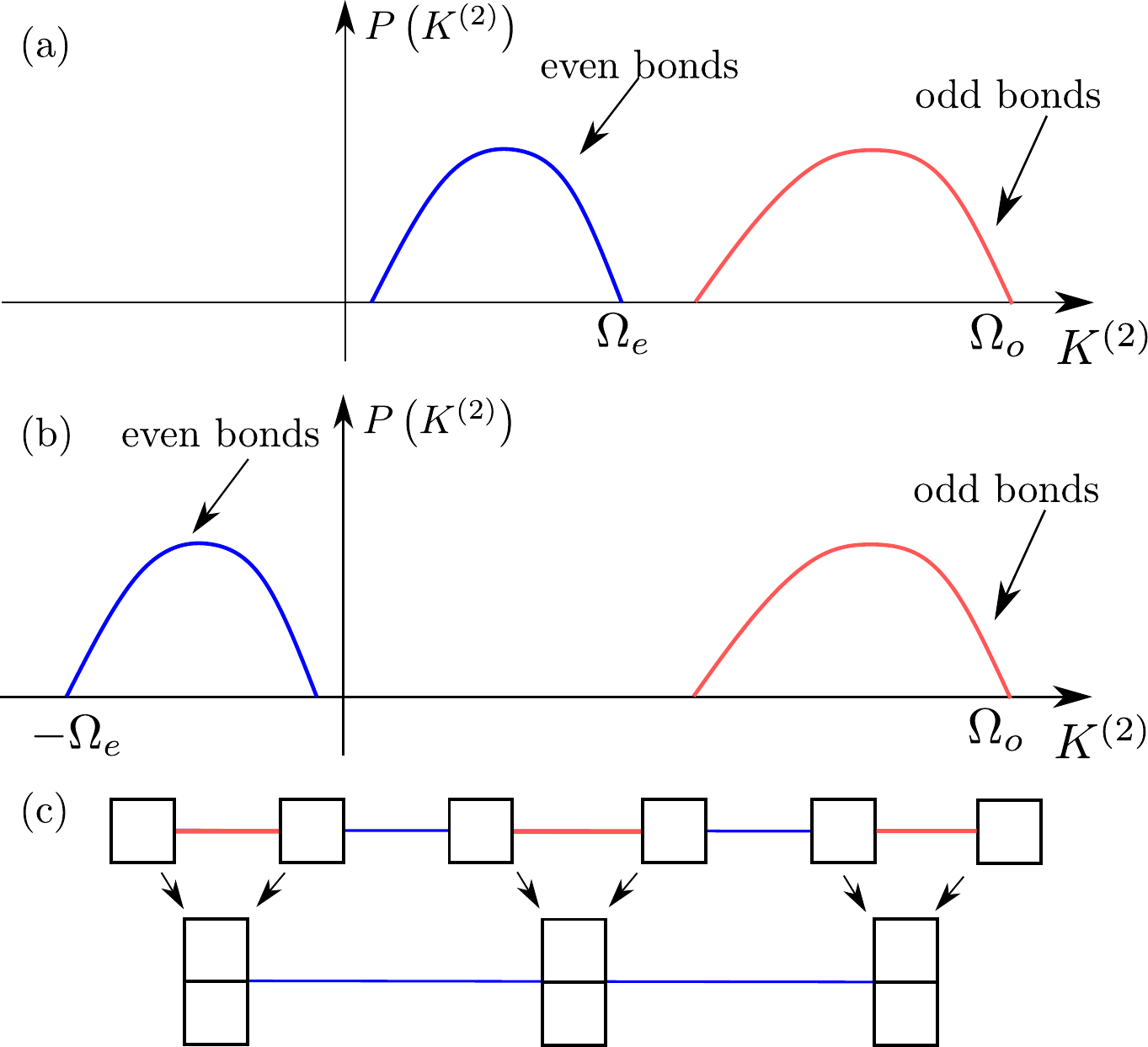}

\caption{An SO(4) symmetric chain without an emergent SU(4) symmetry. (a) and
(b) Initial probability distributions of even (blue) and odd (red)
bonds. In (a), the even bonds have $K_{i,e}^{\left(2\right)}>0$,
which corresponds to the case with an emergent SU(4) symmetry. In
(b), $K_{i,e}^{\left(2\right)}<0$, the case where there is no symmetry
enhancement. (c) Evolution of the non-decimated sites as the RG scale
$\Omega$ runs in the range $\Omega_{e}\lesssim\Omega\lesssim\Omega_{o}$.
The odd-bond distribution has a cutoff $\Omega_{o}\gg\Omega_{e}$
such that these bonds will all be decimated first, generating 6-dimensional
representations connected by blue bonds. \label{fig:Contrast-emergence}}
\end{figure}

Consider a \emph{dimerized} SO(4) chain governed by the Hamiltonian
of Eq.~(\ref{eq:HamiltSON}) with $K_{i}^{\left(1\right)}=0$ for
every $i$ (this restriction can be relaxed but it makes the argument
more transparent). The values of the $K_{i}^{\left(2\right)}$ couplings
are random and depend on whether they are on odd, $K_{i,o}^{\left(2\right)}\equiv K_{2i+1}^{\left(2\right)}$,
or even, $K_{i,e}^{\left(2\right)}\equiv K_{2i}^{\left(2\right)}$,
bonds. The odd couplings are taken to be strictly positive and much
larger than the absolute values of the couplings on even bonds $K_{i,o}^{\left(2\right)}\gg\left|K_{i,e}^{\left(2\right)}\right|$.
The even couplings can be either \emph{all positive} or \emph{all
negative}. This is shown schematically in Fig.~\ref{fig:Contrast-emergence}(a)
and (b), respectively.

We now refer to the ground multiplet structure of a pair of spins,
as shown in Fig.~\ref{fig:levels-so4}. In the initial stages of
the RG, only odd bonds will be decimated and this will give rise to
effective spins transforming as the 6-dimensional representation of
SO(4), see the innermost arc of Fig.~\ref{fig:levels-so4}. Since
these decimations are performed in first order of perturbation theory,
the distribution of even bonds will not change appreciably. Moreover,
the signs of the even bonds will remain the same: either all positive
or all negative, as given by Eq.(\ref{eq:firstorderdecimation2})
with $\xi_{1,3}=1$ and $\Phi_{1,3}=1/2$. After all the odd bonds
have been decimated, we will be left with an \emph{effective chain
of spins belonging to the 6-dimensional representation of SO(4)} with
random $K_{i,e}^{\left(2\right)}$ couplings only, either all positive
or all negative. This is shown schematically in Fig.~\ref{fig:Contrast-emergence}(c)

The next decimations will be governed by the outermost arc of Fig.~\ref{fig:levels-so4}.
At each decimation, the ground state is always a singlet. From Eq.~(\ref{eq:K-order2-2}),
the decimation rule, particularized to case of $N=4$ and $Q=2$ (the
6-dimensional representation), is

\begin{equation}
\tilde{K}_{1,4\,,e}^{\left(2\right)}=\frac{1}{3}\frac{K_{1,e}^{\left(2\right)}K_{3,e}^{\left(2\right)}}{K_{2,e}^{\left(2\right)}}.
\end{equation}
Recall that here, sites 2 and 3 are removed from the chain and an
effective coupling between sites 1 and 4 is generated. Note also that
the decimation rule is valid for any sign of the couplings and thus
\emph{preserves the signs of the initial distribution}.

Even though the ground state is always a singlet, two possible kinds
of singlets can be formed, depending on the sign of $K_{i,e}^{\left(2\right)}$.
Remarkably, only the singlet formed when $K_{i,e}^{\left(2\right)}>0$
is also an SU(4) singlet (the continuous blue line of the the outermost
arc of Fig.~\ref{fig:levels-so4}). When $K_{i,e}^{\left(2\right)}<0$,
by contrast, the singlet \emph{is not} an SU(4) singlet (the dashed
blue line of the the outermost arc of Fig.~\ref{fig:levels-so4}).
As a result, while the former situation exhibits an emergent SU(4)
symmetry, as described in the previous Section~\ref{sec:non-linear-suscep},
the latter one does not. This will be clearly reflected in the physical
properties as we will now show.

The crucial impact of having $K_{i,e}^{\left(2\right)}<0$ is that
the contributions from the singlets no longer vanish in the linear
susceptibilities of the \emph{tensor operators}, see Eq.~(\ref{eq:linear_suscep}).
Physically, an external field coupled to the tensor operators \emph{is
able to polarize the singlets}, as opposed to the effect of a field
coupled to the SO(N) generators. Explicitly, if the singlets are formed
by coupling the 6-dimensional spins at sites $i$ and $i+1$, the
expectation value of tensor operators can be calculated for SO(4)
to be

\begin{equation}
\left\langle \left(\Lambda_{i}^{\beta}+\Lambda_{i+1}^{\beta}\right)^{2}\right\rangle _{\text{singlet}}=\frac{8}{3}.
\end{equation}
Thus, using $n\left(T\right)\ll1$, (\ref{eq:linear_suscep}) now
gives the leading low-temperature behavior

\begin{equation}
\chi_{\beta}^{\left(1\right)}\sim\frac{1}{T}.
\end{equation}
Even though the singlets in this case are not SU(4) singlets they
are obviously still SO(4) singlets. Therefore, their contributions
to the non-linear susceptibilities of the generators remain zero,
and Eq.~(\ref{eq:non-linear-suscep2}) is still valid
\begin{equation}
T^{3}\chi_{\alpha}^{\left(3\right)}+3T^{2}\left[\chi_{\alpha}^{\left(1\right)}\right]^{2}\sim\frac{1}{\left|\log T\right|^{1/\psi}}\underbrace{\neq}_{\text{no\ ES}}T\chi_{\beta}^{\left(1\right)},
\end{equation}
where ``$\mathrm{no\ ES}$'' means that we are dealing with a situation
in which there is no emergent symmetry. We thus see that the connection
between the non-linear susceptibilities of vector operators and the
linear susceptibilities of tensor ones cannot be established in this
case. Other counterexamples can be constructed for other even values
of N.

\section{Discussion and conclusions \label{sec:conclusion}}

\subsection{General results}

We have determined the ground-state structure and the low-temperature
thermodynamic properties of disordered spin chains invariant under
transformations of a large class of Lie groups {[}SO($N$), Sp($N$){]}
in the strong disorder limit. We have determined the phase diagram
and fully characterized the phases when the spins belong to the totally
anti-symmetric representations of the groups (which include the fundamental
one). When the chains have special orthogonal SO($N$) or symplectic
Sp($N$) symmetries at the microscopic level, these phases share the
same physics of chains which are symmetric under transformations of
the larger special unitary SU($N$) group, even though the microscopic
fixed-point Hamiltonian does not have such enlarged symmetry. This
is the defining characteristic of a system exhibiting symmetry emergence,
exposed here by the way the ground state and low-energy excitations
transform under SU($N$) rotations.

Two distinct phases are found, both of them are critical and governed
by infinite-randomness fixed points. The transition between them is
also governed by an infinite-randomness fixed point, albeit with exact
SU($N$) symmetry. Thus, our methods are asymptotically exact in their
fixed-point basins of attraction. The ground states in both phases
and at the transition point are composed of SU($N$) singlets and
thus these phases are of the random-singlet type. The distinction
between these phases stems from the structure of the singlets. In
the meson-like phase, the singlets are composed by only two spins
whereas in the baryon-like phase {[}which occurs only for the SO($N>2)$
cases{]}, the singlets are composed by multiples of $N$ spins.

As shown in Sec.~\ref{sec:Emergent-Symmetries}, two critical and
disorder-independent exponents describe the asymptotic behavior of
these critical phases: the tunneling exponent $\psi$ ($\psi_{M}=\frac{1}{2}$
and $\psi_{B}=\frac{1}{N}$ in the mesonic and baryonic phases, respectively),
which governs the low-temperature thermodynamics, the typical value
of the ground-state correlations, decaying as an stretched exponential,
and the phase-independent $\eta=2$ exponent which governs the mean
value of the correlations {[}that decays as a power law, as in Eq.~(\ref{eq:mean-C}){]}.
Notice the remarkable difference between the arithmetic and typical
averages of the correlations, a hallmark of the infinite-randomness
character of the random singlet phases. Finally, the transition between
these phases is governed by the baryon-like SU($N$) infinite-randomness
fixed-point.

\subsection{Other infinite-randomness universality classes}

Given the variety of tunneling exponent values $\psi$ here found
and its importance in characterizing the corresponding infinite-randomness
fixed points, it is natural to inquire whether $\psi$ can assume
values different from the inverse of an integer $\frac{1}{N}$ with
$N>1$.

The simple answer is yes. Whenever the disorder in the coupling constants
is long-range correlated~\cite{rieger-igloi-prl99} or deterministic,~\cite{vieira-prl05}
$\psi$ can be even greater than $\frac{1}{2}$. In higher dimensions,
$\psi$ can be different as well due to a nontrivial coordination
number.~\cite{senthil-sachdev-prl96,motrunic-etal-prb00} However,
these systems have additional ingredients not contained in our simpler
model. Therefore, we ask whether ``simple'' random systems (i.e.,
systems in which the fixed-point coordination number is exactly $2$\footnote{For instance, random spin ladders have bare coordination number different
from $2$. However, it was shown that the fixed-point Hamiltonians
is one-dimensional, i.e., the low-energy physics is that of a spin
chain.~\cite{hoyos-miranda-ladders}} with irrelevant short-range correlated disorder) can be governed
by an infinite-randomness fixed points with $\psi^{-1}$ being different
from an integer.

Novel values of $\psi$, different from the more conventional one
of $\frac{1}{2}$,~\cite{fishertransising2,fisher94-xxz} were first
found in certain random Heisenberg spin-$S$ chains with $\psi=\frac{1}{m}$,
with $m=2S+1$ being the number of different dimerized phases meeting
at the multicritical point.~\cite{refael-s32,Damle2002,PhysRevB.66.104425}
The corresponding universality class was named permutation symmetric
due to the $m$ distinct domains coexisting at the multicritical fixed
point. However, the corresponding Hamiltonian required to ensure that
all those phases meet at the same point is not known for $m>4$.~\footnote{The possibility of fine tuning raised by the authors was the addition
of next-nearest neighbor interactions. However, this seems unlikely
in view of the results of Ref.~\onlinecite{hoyos-miranda-ladders}.}

The first concrete model realizing the permutation-symmetric infinite-randomness
universality class for $m>4$ and meeting our criterion of being a
``simple system'' was the random anti-ferromagnetic SU($N$)-symmetric
spin chain.~\cite{PhysRevB.70.180401} It was shown to be governed
by the baryonic infinite-randomness fixed points with $\psi=\psi_{B}=\frac{1}{N}$.
Later, the SU(2)$_{k}$-symmetric anyonic chains~\cite{fidkowski-etal-prb09}
(with $k$ being an odd integer) was the second realization of this
universality class where $\psi=\frac{1}{k}$. Finally, the random
SO($N$)-symmetric spin chains here studied constitute yet another
example. In all these systems, the SDRG decimation rules can be mapped
into each other. The topological charge carried by the anyons plays
the role of the domain-wall spins which, in turn, play the role of
our totally antisymmetric spin representations in the SO($N$) language.
The corresponding sign of the couplings (ferro- or anti-ferromagnetic)
between the anyons or between the domain-wall spins play the role
of the two angles found in our baryonic fixed points. In all cases,
the tunneling exponent is determined by the probability $p$ of having
certain link configurations (ensuring a second-order decimation into
a singlet). Thus, $p=\frac{N_{2}}{N_{T}}$ where $N_{2}$ is the total
number of link configurations ensuring a singlet decimation and $N_{T}$
is the total number of distinct configurations. From symmetry, these
configurations are all equally probable and thus, $N_{T}$ must be
a multiple (which turns out to be $N-1$) of $N_{2}$. Thus, $p=\frac{1}{N-1}$
is the inverse of an integer implying that the tunneling exponent
$\psi=\frac{p}{1+p}=\frac{1}{N}$ is also the inverse of an integer.
One might expect to find new values of $\psi$ in SU($N$)$_{k}$
anyonic spin chains. However, this would only change $p=\frac{1}{N-1}\rightarrow\frac{1}{\left(k-1\right)\left(N-1\right)}$
which is also the inverse of an integer.

Another place to search for different values of $\psi$ could be in
random spin chains invariant under transformations of a discrete symmetry
group. However, many quantum critical chains of random models like
the Ising, and the various $N$-state Potts, clock, parafermionic
and Ashkin-Teller models were studied.~\cite{fishertransising2,senthil-majumdar-prl96,santachiara-06,hrahsheh-hoyos-vojta-prb12,barghathi-etal-PS15}
In all cases, the fraction of second-order decimations is $p=1$,
and thus $\psi=\frac{1}{2}$. Indeed, the ground state of the quantum
critical Ising chain can be described as an RSP of an SU(2)$_{2}$
random anyonic chain.~\cite{bonesteel-yang-prl07} It is plausible
that the other models may also be described likewise.

Evidently, we do not claim to have exhausted all possible infinite-randomness
universality classes in simple one-dimensional systems. However, given
the plethora of examples mentioned above, it is conceivable that the
permutation-symmetric infinite-randomness universality class is the
most general one capable of producing different values of $\psi$
in one-dimensional systems fulfilling our criterion of ``simple systems''.

\subsection{Final remarks}

The situations analyzed here have particular interest due to the possibilities
of experimental realization,~\cite{QuitoLopes2017PRL} but do not
necessarily exhaust all possibilities of symmetry enhancement in disordered
spin chains. They do provide, however, a very large class of systems
showing this phenomenon. The methodology developed here is also very
embracing and provides the guidelines and tools for the study of more
involved, exotic or simply distinct scenarios. Exceptional Lie algebras
remain to be studied, as well as Hamiltonians starting with larger
dimensional representations of SO($N$) and Sp($N$) at each site.
These problems are left for future research.

\section{acknowledgments}

We thank Gabe Aeppli for discussions. VLQ and PLSL contributed equally
to this work. VLQ acknowledges financial support from the NSF Grant
No. DMR-1555163 and the National High Magnetic Field Laboratory through
DMR-1157490, the State of Florida and the Aspen Center for Physics,
supported by NSF grant PHY-1607611, for hospitality. PLSL is supported
by the Canada First Research Excellence Fund. JAH acknowledges financial
support from FAPESP and CNPq. EM acknowledges financial support from
CNPq (Grant No. 307041/2017-4) and Capes (Grant No. 0899/2018).

\appendix

\section{Derivation of the $\Phi$ and $\xi$ listed in Table~\ref{tab:List-of-beta}
\label{sec:Derivation-xi-Phi}}

In this Appendix, we derive the values of $\Phi$ and $\xi$ listed
in Table~\ref{tab:List-of-beta}. The derivation follows closely
the SU(N) SDRG steps studied in Ref.~\onlinecite{PhysRevB.70.180401}.
We use the notation $\mathcal{Q}$ and $Q$ for the SU(N) and SO(N)
Young tableaux, respectively, as well as $J_{i}$ for the couplings
of the SU(N) chain. The SU(N) decimation rules of Ref.~\onlinecite{PhysRevB.70.180401},
when the ground state is not a singlet, are

\begin{equation}
\tilde{J}_{1}=\begin{cases}
\xi J_{1} & \mathcal{Q}_{2}+\mathcal{Q}_{3}<N\\
\bar{\xi}J_{1} & \mathcal{Q}_{2}+\mathcal{Q}_{3}>N
\end{cases}\label{eq:J1-SUN}
\end{equation}

\begin{equation}
\tilde{J}_{3}=\begin{cases}
\left(1-\xi\right)J_{3} & \mathcal{Q}_{2}+\mathcal{Q}_{3}<N\\
\left(1-\bar{\xi}\right)J_{3} & \mathcal{Q}_{2}+\mathcal{Q}_{3}>N
\end{cases},
\end{equation}
where $\xi=\frac{\mathcal{Q}_{2}}{\mathcal{Q}_{2}+\mathcal{Q}_{3}}$,
$\bar{\xi}=\frac{\mathcal{\bar{Q}}_{2}}{\mathcal{\bar{Q}}_{2}+\mathcal{\bar{Q}}_{3}}$,
and $\bar{\mathcal{Q}}\equiv N-\mathcal{Q}$. 

Building a correspondence between the SO(N) and SU(N) Young tablaeux
and comparing the decimation rules with Eq.~(\ref{eq:J1-SUN}), each
line of Table~\ref{tab:List-of-beta} is fixed. The first line, found
in the positive $K^{\left(2\right)}$ region, is straightforward.
If two SO(N) representations $Q_{2}$ and $Q_{3}$ are added generating
$\tilde{Q}=Q_{2}+Q_{3}\le\text{int}\left(\frac{N}{2}\right)$, the
correspondence with SU(N) representations is immediate, that is, $Q_{2,3}=\mathcal{Q}_{2,3}$
and $\tilde{Q}=$ $\tilde{\mathcal{Q}}$, and the decimation rules
are exactly the same as in Eq.~(\ref{eq:J1-SUN}).

If, on the other hand, $Q_{2}+Q_{3}>\text{int}\left(\frac{N}{2}\right)$
(second line of Table~\ref{tab:List-of-beta}), it follows that $\tilde{\mathcal{Q}}=\mathcal{Q}_{2}+\mathcal{Q}_{3}<N$
in the SU(N) language. When translated to the SO(N) tableaux, $\tilde{Q}=N-Q_{2}-Q_{3}$.
It follows, therefore, that $N-Q_{2}-Q_{3}=\mathcal{Q}_{2}+\mathcal{Q}_{3}$.
This transformation of representations is achieved by choosing $\xi_{1}=\xi_{3}=-1$,
fixing the next entry of Table~\ref{tab:List-of-beta}.

Finally, for the decimations of negative $K^{\left(2\right)}$ (last
line of Table~\ref{tab:List-of-beta}), the SU(N)-SO(N) identifications
$\mathcal{Q}_{2}=N-Q_{2}$ and $\mathcal{Q}_{3}=Q_{3}$ are made.
Notice also that we chose $Q_{2}<Q_{3}$. This correspondence can
be seen by comparing the Hamiltonian at the SU(N)-invariant point
$\theta=-\frac{\pi}{4}$. By performing the RG decimation using the
SU(N) language, the ground state is $\tilde{\mathcal{Q}}=N-Q_{2}+Q_{3}$.
From the choice $Q_{3}>Q_{2}$, $\tilde{\mathcal{Q}}>N$, and within
SU(N), we get $\tilde{\mathcal{Q}}=Q_{3}-Q_{2}$. Putting everything
together, Eq.~(\ref{eq:J1-SUN}) returns

\begin{align}
\tilde{J}_{1} & =\frac{\mathcal{\bar{Q}}_{2}}{\mathcal{\bar{Q}}_{2}+\mathcal{\bar{Q}}_{3}}J_{1}\nonumber \\
 & =\frac{N-\mathcal{Q}_{2}}{2N-\mathcal{Q}_{2}+\mathcal{Q}_{3}}J_{1}\nonumber \\
 & =\frac{Q_{2}}{N-Q_{2}+Q_{3}}J_{1}.
\end{align}
This gives the value of $\Phi_{1}$ listed in the third line of Table~\ref{tab:List-of-beta}.

The derivation of $\Phi_{3}$ follows similar steps.

\bibliographystyle{apsrev4-1}
\bibliography{bibliog_son,all}

\end{document}